\documentclass[10pt,journal]{IEEEtran}
\ifCLASSINFOpdf
\else
\fi
\usepackage[latin1]{inputenc}

\usepackage{amsmath}
\usepackage{amssymb}
\usepackage{graphicx}
\usepackage{hyperref}
\usepackage{amscd}
\usepackage{mathrsfs}
\usepackage{bbm}
\usepackage{threeparttable}
\usepackage{caption}
\usepackage{amsthm}

\usepackage{amscd}
\usepackage{mathrsfs}
\usepackage{bbm}
\usepackage{amsfonts}
\usepackage{epic}
\usepackage[misc]{ifsym}
\usepackage{url}

\usepackage{algorithmic}
\usepackage{comment}
\usepackage{bm}
\usepackage{pifont}
\newcommand{\cmark}{\ding{51}}%
\newcommand{\xmark}{\ding{55}}%

\usepackage{xcolor}
\usepackage{pdfcolmk}
\usepackage{booktabs}
\usepackage{textcomp}
\usepackage{mathrsfs}

\usepackage{enumitem}
\newtheorem{theorem}{Theorem}

\theoremstyle{definition}

\theoremstyle{remark}
\newtheorem{remark}[theorem]{Remark}
\renewcommand \thetable{\arabic{table}}
\usepackage{array}
\usepackage{booktabs}

\usepackage[ruled]{algorithm2e}
\usepackage{subfigure}
\usepackage{multirow}

\begin{document}
%
\title{FSSA: Efficient 3-Round Secure Aggregation for Privacy-Preserving Federated Learning}
%
%
%

\author{Fucai~Luo, Saif~Al-Kuwari, Haiyan Wang, and Xingfu Yan 
\thanks{This work was supported in part by the Guangxi Natural Science Foundation (No. 2022GXNSFBA035650) and the Major Key Project of PCL (No. PCL2022A03).  (Corresponding author: Haiyan Wang and Xingfu Yan) \newline
\indent Fucai Luo is with the College of Computer and Information Engineering, Zhejiang Gongshang University, Hangzhou, China.  E-mail: lfucai@126.com. \newline
\indent Saif Al-Kuwari is with the College of Science and Engineering, Hamad Bin Khalifa University, Doha, Qatar.  E-mail: smalkuwari@hbku.edu.qa. \newline
\indent Haiyan Wang is with the Department of New Networks, Peng Cheng Laboratory, Shenzhen, China.  E-mail: wanghy01@pcl.ac.cn. \newline
\indent Xingfu Yan is with the School of Computer Science, South China Normal University, Guangzhou, China. E-mail: xfyan78@163.com.

}

}

\maketitle

\begin{abstract}
Federated learning (FL) allows a large number of clients to collaboratively train machine learning (ML) models by sending only their local gradients to a central server for aggregation in each training iteration, without sending their raw training data.  Unfortunately, recent attacks on FL demonstrate that local gradients may leak information about local training data.  In response to such attacks, Bonawitz \textit{et al.} (CCS 2017) proposed a secure aggregation protocol that allows a server to compute the sum of clients' local gradients in a secure manner.   However, their secure aggregation protocol requires at least 4 rounds of communication between each client and the server in each training iteration.  The number of communication rounds is closely related not only to the total communication cost but also the ML model accuracy, as the number of communication rounds affects client dropouts.

In this paper, we propose FSSA, a 3-round secure aggregation protocol, that is efficient in terms of computation and communication, and resilient to client dropouts.   We prove the security of FSSA in honest-but-curious setting and show that the security can be maintained even if an arbitrarily chosen subset of clients drop out at any time.  We evaluate the performance of FSSA and show that its computation and communication overhead remains low even on large datasets.  Furthermore, we conduct an experimental comparison between FSSA and Bonawitz \textit{et al.}'s protocol.  The comparison results show that, in addition to reducing the number of communication rounds,  FSSA achieves a significant improvement in computational efficiency.
\end{abstract}


\IEEEpeerreviewmaketitle

\section{Introduction}
Federated learning (FL) \cite{KMY16,lf22} is a promising collaborative machine learning (ML) framework allowing a large number of participating entities
to collaboratively train ML models under the orchestration of a server.  In particular, FL enables clients to locally train ML models on their data independently and in parallel, which greatly reduces data privacy risks and improves training efficiency and scalability.

In each iteration of FL, the server first selects a subset of clients and sends them a copy of the current global model.  Next, each client trains the model locally on their data and computes a local gradient, which is then sent to the server.  Finally, the server aggregates these gradients, updates the global model and proceeds to the next training iteration.  However, recent results \cite{PAH18,SSSS17,GWYGB18} showed that clients' local gradients may leak information about the corresponding local training data.  In addition, these participating clients are usually a large number of mobile or edge devices (e.g., smartphones, personal computers, and IoT devices) with unreliable communication and limited computing power, and therefore can drop out of the protocol. Consequently, the FL system must be resilient to client dropouts.  In particular, since the client may drop out of the protocol at any time, the fewer the number of interactions between the server and the client, the lower the number of client dropouts (i.e., client dropout rate).  Thus, the number of communication rounds not only affects the total communication cost but also affects the client dropout rate, which in turn affects the model accuracy.  Given these concerns,  the FL protocol should be optimized to protect the privacy of clients' local gradients while being highly communication efficient as well as being resilient to client dropouts.

Bonawitz \textit{et al.} \cite{BIK17} proposed the first secure aggregation protocol to address the privacy of local gradient and client dropouts.  At a high level, secure aggregation is a secure multiparty computation (MPC) that allows the server to compute the sum of the clients' local gradients without having to reveal the clients' local gradients (even to the server).   In fact,  secure aggregation inspired several different approaches including: secure multiparty computation (MPC) \cite{BCP15}, partially\slash fully homomorphic encryption \cite{Pa99,MW16}, functional encryption \cite{AMDD18}, and double-masking \cite{BIK17,CSH20}.  However, most of these solutions either require more communication rounds \cite{BIK17,so21},
do not tolerate client dropouts \cite{T19,BCT20}, use heavy cryptographic primitives \cite{DYC20,sa20}, and/or rely on a trusted third party \cite{HML19,ZWR20}.   In this paper, we focus on the secure aggregation protocols using double-masking.


\subsection{Contributions}

In this paper, we propose FSSA, a 3-round secure aggregation protocol, that is efficient in terms of computation and communication, and resilient to client dropouts.  Technically, FSSA is based on the secure aggregation framework of Bonawitz \textit{et al.} \cite{BIK17} and uses ramp secret sharing \cite{BM84}, key agreement \cite{di76},  and symmetric encryption to: 1) protect the privacy of the local gradient and,  2) handle client dropouts.  Compared with \cite{BIK17}, FSSA removes the use of double-masking technique, thereby reducing one round of communication.
Our contributions can be summarized as follows:
\begin{itemize}
  \item We exploit an efficient $(t,d,n)$-ramp secret sharing, in conjunction with key agreement and symmetric encryption to achieve secure aggregation and tolerance to client dropouts.

  \item We provide comprehensive security analysis for FSSA, using a simulation-based proof, which is a standard technique for security analysis of MPC protocols.  We also evaluate the performance of FSSA and show that its computation and communication overhead remains low even on large datasets.

  \item We give an experimental comparison between FSSA and Bonawitz \textit{et al.}'s protocol \cite{BIK17}.  The comparison results show that, compared with \cite{BIK17}, FSSA has a higher computational cost per client, a lower computational cost on the server, and a lower total computational cost.  This demonstrates that in addition to reducing the number of communication rounds, FSSA achieves a significant improvement in computational efficiency.  However, in terms of overall communication overhead, FSSA does not perform as well as the secure aggregation protocol in \cite{BIK17}, which can be seen as a trade-off for getting fewer communication rounds and lower computational overhead.
\end{itemize}

To our knowledge, the secure aggregation protocols using double-masking technique proposed by \cite{BIK17} are more efficient than other protocols (e.g. those based on homomorphic encryption or MPC), while the few other protocols that require fewer communication rounds either rely on expensive cryptographic primitives (e.g. homomorphic encryption and multi-input functional encryption), or require multiple servers (this usually requires the assumption that there is no collusion between servers; as opposed to the single server required by the mainstream solutions) and do not tolerate client dropouts.  These can be seen from Table \ref{T1} and related work in Section \ref{AA1}.   For those based on double-masking technique (e.g. \cite{CSH20,b20,guo20v,XLL20}), their core is still the double-masking technique for secure aggregation (see related work in Section \ref{AA1} for more details).  In conclusion, we believe that a comparison with \cite{BIK17} is sufficient to illustrate the advantages and disadvantages of our FSSA, especially since in this work we are only concerned with reducing the number of communication rounds for secure aggregation protocols while maintaining efficiency.

\begin{remark}
As clients in FL join the secure aggregation protocol at different points in time, in any given time interval (for example, in a few minutes, which is much longer than the running time of each client) the server must wait until the last client joins the protocol and then performs an aggregation operation.  This means that the execution time of the protocol is mainly determined by:  1) the time of waiting for clients to join the protocol and, 2) the running time of the server (here, we assume that each client has already trained the model locally before joining the protocol).  In other words, the running time of the server has a greater impact on the training speed of the model than the running time of each client.
\end{remark}

\subsection{Organization}
The remainder of the paper is organized as follows.   We briefly outline related work in Section \ref{AA1}.  In Section \ref{Sec:2}, we review some cryptographic primitives we use in our protocol.   We then give a technical overview of our protocol in Section \ref{Sec:3}, followed by a formal protocol description in Section \ref{Sec:4}.   In Section \ref{Sec:5}, we give the correctness and security analysis of our protocol.   We evaluate the performance of the proposed protocol and compare it with the previous secure aggregation protocol in Section \ref{Sec:6}.   Finally,  Section \ref{Sec:7} concludes the paper.
\begin{table*}
\centering
\caption{Comparison of privacy-preserving approaches in FL framework.}
\vspace{0.2cm}
\newcommand{\tabincell}[2]{\begin{tabular}{@{}#1@{}}#2\end{tabular}}
\begin{threeparttable}
\begin{tabular}{ccccccccccccc}
\toprule[1pt]
Approach &No Trusted Third Party &No Expensive Operations&Resilient Against Dropouts& Rounds\\
\specialrule{0em}{1.5pt}{1.5pt}
       \hline
       \specialrule{0em}{2pt}{2pt}
Secure Aggregation with Secret Sharing &\cmark
 & \cmark & \cmark & $\geq 4$ \\
 \cite{BIK17,CSH20,b20,guo20v,XLL20} &
 &  &  &\\
 \specialrule{0em}{3pt}{3pt}
 Turbo-Aggregate \cite{so21}&\cmark & \cmark & \cmark & $O(n/\log n)$ \\
\specialrule{0em}{3pt}{3pt}
SAFER \cite{BCT20} &\cmark & \cmark & \xmark & 2\tnote{$\dagger$} \\
\specialrule{0em}{3pt}{3pt}
Secure Aggregation with HE
&\xmark
 & \xmark & \cmark \tnote{$\ddagger$} & $\geq 2$\tnote{$\ddagger$}  \\
 \cite{PAH18,ZWR20,HML19,T19,DYC20,ZLX20} &
 &  &  &\\
\specialrule{0em}{2pt}{2pt}
SAFELearn \cite{FHM21} &\cmark& \cmark & \cmark&  2\tnote{$\dagger$}  \\
\specialrule{0em}{2pt}{2pt}
 POSEIDON \cite{sa20} &\cmark& \xmark &\cmark \tnote{$\star$} &  $\geq2$\tnote{$\ast$}  \\
\specialrule{0em}{2pt}{2pt}
Hybridalpha \cite{xu19} &\xmark& \xmark & \cmark&  2\tnote{$\ddagger$}  \\
\specialrule{0em}{2pt}{2pt}
FSSA (This work) &\cmark& \cmark & \cmark  & 3 \\
  \bottomrule[1pt]
\end{tabular}\vspace{0.1cm}
\begin{tablenotes}
\item [$\dagger$] It requires multiple non-colluding servers.
\item [$\ddagger$] When considering the key distribution communication round.  In addition, some of these approaches using threshold-based homomorphic cryptosystems do not tolerate client dropouts and require an extra round of communication.
\item [$\star$] When POSEIDON does not use the decentralized bootstrapping.
\item [$\ast$] The distributed bootstrapping requires an extra round of communication.
\end{tablenotes}
\end{threeparttable}
\label{T1}
\end{table*}

\section{Related Work}
\label{AA1}
In this section, we survey the existing secure aggregation protocols based on different approaches, which are summarized in Table \ref{T1}. \\ 

\noindent \textbf{Secure Aggregation with Secret Sharing.} The first secure aggregation protocol was proposed by Bonawitz \textit{et al.} \cite{BIK17}, who used double-masking technique (with pseudorandom values),  Shamir's secret sharing \cite{sh79}, key agreement \cite{di76},  and symmetric encryption to protect the privacy of the local gradients and handle client dropouts.   However, their secure aggregation protocol requires at least $4$ rounds of communication between each client and the server in every training iteration.

Based on the framework of \cite{BIK17}, Bell \textit{et al.} \cite{b20} and Choi \textit{et al.} \cite{CSH20} proposed secure aggregation protocols with polylogarithmic communication and computation overhead.  Their protocols achieve better computational and communication efficiency than the secure aggregation protocol of \cite{BIK17} but they are both much more involved and do not reduce the number of communication rounds.   The key idea behind \cite{b20} and \cite{CSH20} is to replace the complete communication graph of \cite{BIK17} by a sparse random graph and to use secret sharing only for a subset of clients instead of for all clients.   However, to ensure the correctness and security of their protocols, the sparse random graph must satisfy a sequence of rigorous conditions (e.g., not too many corrupt neighbors, connectivity after dropouts, not too many neighbors drop out), which make their protocols not suitable for practical applications.  In addition, these approaches are generic and therefore apply to our FSSA, but in this work we only focus on how to reduce the number of communication rounds of secure aggregation protocols.

To address efficiency challenge, So \textit{et al.} \cite{so21} recently proposed Turbo-Aggregate, which uses a circular communication topology to reduce the computation and communication overhead of \cite{BIK17}.  However, Turbo-Aggregated requires $O(n/\log n)$ rounds of communication, where $n$ is the number of clients.  Beguier \textit{et al.} \cite{BCT20} proposed SAFER, a secure aggregation protocol between multiple servers with low computation and communication overhead.  SAFER achieves low computational and communication costs by combining update compression technique with arithmetic sharing.  However, SAFER requires multiple servers and does not tolerate client dropouts.


On the other hand, to guarantee the correctness of the aggregated gradient provided by the server,  based on the secure aggregation framework of \cite{BIK17}, Xu \textit{et al.}  and Guo \textit{et al.} proposed VerifyNet \cite{XLL20} and VeriFL \cite{guo20v} respectively.  Technically, VerifyNet and VeriFL add verifiability of aggregated gradient on top of \cite{BIK17}, so that each client can verify the correctness of the aggregated gradient, thus ensuring that the server computes the sum of clients' local gradients honestly.  However, these verifiable secure aggregation protocols only achieve verifiability (even at the cost of more communication rounds) without improving the efficiency of secure aggregation.\\

\noindent \textbf{Secure Aggregation with Encryption.}  Homomorphic Encryption (HE) allows certain operations (e.g., addition) to be performed directly on encrypted data.  Such property is exactly what is needed for secure aggregation.  A number of proposals \cite{PAH18,ZWR20,HML19,T19,DYC20,ZLX20,LYJ20,CWC23} have been made to build a secure aggregation protocol for FL using additively HE.

Recently, Fereidooni \textit{et al.} \cite{FHM21} proposed SAFELearn, a generic design for secure (private) aggregation.  SAFELearn can be instantiated with MPC or HE.  However, unlike mainstream secure aggregation protocols, SAFELearn requires two servers and assumes no collusion between the two servers.  Sav \textit{et al.} \cite{sa20} proposed POSEIDON, using a multiparty lattice-based homomorphic encryption scheme \cite{MCT20}.  To improve the efficiency of POSEIDON, the authors provided a generic packing approach so that single-instruction-multiple-data (SIMD) operations can be efficiently performed on encrypted data.  However, POSEIDON only supports client dropouts when the decentralized bootstrapping is not used.  Similarly, Xu \textit{et al.} \cite{xu19} proposed Hybridalpha, using a multi-input functional encryption scheme \cite{AMDD18} and differential privacy (DP) \cite{DC14}.   With multi-input functional encryption, each client obtains a public key and uses it to encrypt the local gradient, while the server obtains a function key and uses it to compute the average cumulative sum of the clients' gradients.  However, most of these secure aggregation protocols rely on expensive cryptographic primitives and a trusted third party (TTP) that generates public\slash private key pairs for all clients.\\

\noindent \textbf{MPC and\slash or DP.}  Other works \cite{MPR18,BRD16,FSS20,pat20,XSW22} combine MPC techniques and\slash or DP with ML.   However, these protocols usually use some heavy cryptographic primitives and are usually customized for specific ML algorithms, which limits their flexibility and scalability.  Therefore, these types of approaches are not suitable for FL.



\section{Preliminaries}
\label{Sec:2}
In this section, we review some cryptographic primitives used in our protocol.  In what follows, we let $[n]\triangleq\{1,\ldots,n\}$ for any positive integer $n$, and $\lambda$ be the security parameter.

\subsection{Secret Sharing}
\label{Sec:2.2}
A secret sharing scheme contains polynomial-time algorithms $\textbf{SS} =(\textbf{SS.Setup}, \textbf{SS.Share}, \textbf{SS.Recon})$, defined as follows:
\begin{itemize}
  \item $\textnormal{\textbf{SS.Setup}}(1^\lambda)\rightarrow \textsf{SSpp}$: outputs a public parameter $\textsf{SSpp}$, which includes a message space $\mathcal{M}$ (e.g., a finite field $\mathbb{F}$).
  \item $\textnormal{\textbf{SS.Share}}(t, \mathcal{U}, s)\rightarrow \{[\![s]\!]_u\}_{u\in \mathcal{U}}\in \mathcal{M}$: outputs $|\mathcal{U}|$ shares of the secret $s\in \mathcal{M}$, where $\mathcal{U}\subseteq \mathcal{M}$ is a set and $1 \leq t\leq |\mathcal{U}|$ is the threshold value.
  \item $\textnormal{\textbf{SS.Recon}}(t, \{[\![s]\!]_u\}_{u\in \mathcal{V}})\rightarrow s$: outputs $s\in \mathcal{M}$, where $\mathcal{V}\subseteq \mathcal{U}$ and $|\mathcal{V}|\geq t$.
\end{itemize}

The above secret sharing scheme satisfies the following requirements:
\begin{enumerate}
  \item \textbf{Correctness.} For any secret $s\in \mathcal{M}$, any set $\mathcal{U}\subseteq \mathcal{M}$ and any threshold $1 \leq t\leq |\mathcal{U}|$, if $\{[\![s]\!]_u\}_{u\in \mathcal{U}} \leftarrow \textnormal{\textbf{SS.Share}}(t, \mathcal{U}, s)$, we have $\textnormal{\textbf{SS.Recon}}(t, \{[\![s]\!]_u\}_{u\in \mathcal{V}})=s$ for any subset $\mathcal{V}\subseteq \mathcal{U}$ satisfying $|\mathcal{V}|\geq t$.
  \item \textbf{Shannon Perfect Security.} For any secrets $s,s'\in \mathcal{M}$, any threshold $1 \leq t\leq |\mathcal{U}|$, any subset $\mathcal{V}\subseteq \mathcal{U}\subseteq \mathcal{M}$ such that $|\mathcal{V}|< t$, we have
\begin{eqnarray*}
  &\{\{[\![s]\!]_u\}_{u\in \mathcal{U}} \leftarrow \textnormal{\textbf{SS.Share}}(t, \mathcal{U}, s): \{[\![s]\!]_u\}_{u\in \mathcal{V}}\} \equiv  \\
   & \{\{[\![s']\!]_u\}_{u\in \mathcal{U}} \leftarrow \textnormal{\textbf{SS.Share}}(t, \mathcal{U}, s'): \{[\![s']\!]_u\}_{u\in \mathcal{V}}\},
\end{eqnarray*}
where $\equiv$ indicates that the two distributions are identical.
  \item \textbf{Linearity.} For any secrets $s_1,s_2\in \mathcal{M}$, given the shares $\{[\![s_1]\!]_u\}_{u\in \mathcal{U}}$ of the secret $s_1$ and shares $\{[\![s_2]\!]_u\}_{u\in \mathcal{U}}$ of the secret $s_2$, then  $\{a[\![s_1]\!]_u+b[\![s_2]\!]_u\}_{u\in \mathcal{U}}$ is shares of the secret $as_1+bs_2$ for any $a,b\in\mathcal{M}$.
\end{enumerate}

A well-known secret sharing scheme is Shamir's secret sharing \cite{sh79}.  However, the coding efficiency of such $(t,n)$-secret sharing scheme is low because the size of secret is less than or equal to the size of shares (i.e., only one secret can be recovered per reconstruction).  This somewhat limits the utility of the original Shamir's secret sharing as many practical applications require high efficiency, even more than Shannon perfect security.    To address this efficiency problem, ramp secret sharing schemes have been proposed \cite{BM84,CCG07,Ta13,ku09fast}, which have a trade-off between efficiency and security.  Not surprisingly, ramp secret sharing schemes have been widely used in numerous applications in cryptography, including secure multiparty computation (MPC) \cite{FY92}, error decodable secret sharing \cite{MPS11} and broadcast encryption \cite{SW99}.

A $(t,d,n)$-ramp secret sharing scheme consists of polynomial-time algorithms $\textbf{RSS} =(\textbf{RSS.Setup}, \textbf{RSS.Share}, \textbf{RSS.Recon})$, described as follows:

 \begin{itemize}
  \item $\textnormal{\textbf{RSS.Setup}}(1^\lambda)\rightarrow \textsf{RSSpp}$: outputs a public parameter $\textsf{RSSpp}$, which includes a message space $\mathcal{M}$.
  \item $\textnormal{\textbf{RSS.Share}}(t, d, \mathcal{U}, \mathbf{s})\rightarrow \{[\![\mathbf{s}]\!]_u\}_{u\in \mathcal{U}} \in \mathcal{M}^{d}$: outputs $|\mathcal{U}|$ shares of the secret $\mathbf{s}\in \mathcal{M}^d$, where $1 \leq t\leq |\mathcal{U}|$ is the threshold value, $d$ is the size of the secret, $\mathcal{U}\subseteq \mathcal{M}$ is a set and $d>0$.
  \item $\textnormal{\textbf{RSS.Recon}}(t, d, \{[\![\mathbf{s}]\!]_u\}_{u\in \mathcal{V}})\rightarrow \mathbf{s}$: outputs $\mathbf{s}\in \mathcal{M}^d$, where $\mathcal{V}\subseteq \mathcal{U}$ and $|\mathcal{V}|\geq t$.
\end{itemize}
The above $(t,d,n)$-ramp secret sharing scheme satisfies the following requirements:
\begin{enumerate}
  \item \textbf{Correctness.} For any secret $\mathbf{s}\in \mathcal{M}^d$, any threshold $1 \leq t\leq n$, any integer $t > d> 0$, and any set $\mathcal{U}\subseteq \mathcal{M}$ with $|\mathcal{U}|=n$, if $\{[\![\mathbf{s}]\!]_u\}_{u\in \mathcal{U}} \leftarrow \textnormal{\textbf{RSS.Share}}(t, d,\mathcal{U}, \mathbf{s})$, we have $\textnormal{\textbf{RSS.Recon}}(t, d, \{[\![\mathbf{s}]\!]_u\}_{u\in \mathcal{V}})=\mathbf{s}$ for any subset $\mathcal{V}\subseteq \mathcal{U}$ satisfying $|\mathcal{V}|\geq t$.
  \item \textbf{Security.} \begin{itemize}
                             \item \textbf{Shannon Perfect Security.}
  For any secrets $\mathbf{s},\mathbf{s}'\in \mathcal{M}^d$, any threshold $1 \leq t\leq n$, any subset $\mathcal{V}\subseteq \mathcal{U}\subseteq \mathcal{M}$ such that $|\mathcal{V}|\leq t-d$, we have
\begin{eqnarray*}
  &\{\{[\![\mathbf{s}]\!]_u\}_{u\in \mathcal{U}} \leftarrow \textnormal{\textbf{RSS.Share}}(t, d, \mathcal{U}, \mathbf{s}): \{[\![\mathbf{s}]\!]_u\}_{u\in \mathcal{V}}\} \equiv  \\
   & \{\{[\![\mathbf{s}']\!]_u\}_{u\in \mathcal{U}} \leftarrow \textnormal{\textbf{RSS.Share}}(t, d, \mathcal{U}, \mathbf{s}'): \{[\![\mathbf{s}']\!]_u\}_{u\in \mathcal{V}}\}.
\end{eqnarray*}

  \item \textbf{$(t-d)$-Privacy.} For any secret $\mathbf{s}\in \mathcal{M}^d$, any threshold $1 \leq t\leq n$, any subset $\mathcal{V}\subseteq \mathcal{U}\subseteq \mathcal{M}$ such that $t-d <|\mathcal{V}|< t$, if $\{[\![\mathbf{s}]\!]_u\}_{u\in \mathcal{U}} \leftarrow \textnormal{\textbf{RSS.Share}}(t, d,\mathcal{U}, \mathbf{s})$, we have that
      the shares $\{[\![\mathbf{s}]\!]_u\}_{u\in \mathcal{V}}$ would leak information about $\mathbf{s}$. (Note that the amount of information leaked depends on the specific scheme)

\end{itemize}

  \item \textbf{Linearity.} For any secrets $\mathbf{s}_1,\mathbf{s}_2\in \mathcal{M}^d$, given the shares $\{[\![\mathbf{s}_1]\!]_u\}_{u\in \mathcal{U}}$ of the secret $\mathbf{s}_1$ and shares $\{[\![\mathbf{s}_2]\!]_u\}_{u\in \mathcal{U}}$ of the secret $\mathbf{s}_2$, then  $\{a[\![\mathbf{s}_1]\!]_u+b[\![\mathbf{s}_2]\!]_u\}_{u\in \mathcal{U}}$ is shares of the secret $a\mathbf{s}_1+b\mathbf{s}_2$ for any $a,b\in\mathcal{M}$.
\end{enumerate}

\subsection{Key Agreement}
A key agreement protocol contains polynomial-time algorithms $\textbf{KA} =(\textbf{KA.Setup}, \textbf{KA.Gen}, \textbf{KA.Agree})$, defined as follows:
\begin{itemize}
  \item $\textnormal{\textbf{KA.Setup}}(1^\lambda)\rightarrow \textsf{KApp}$: outputs a public parameter $\textsf{KApp}$.
  \item $\textnormal{\textbf{KA.Gen}}(\textsf{KApp})\rightarrow (pk_u,sk_u)$: outputs a public\slash secret key pair $(pk_u,sk_u)$ for any client $u$.
  \item $\textnormal{\textbf{KA.Agree}}(sk_u,pk_v)\rightarrow s_{u,v}$: outputs a shared secret key $s_{u,v}$.
\end{itemize}
The above key agreement protocol satisfies the following requirements:
\begin{enumerate}
  \item \textbf{Correctness.} For any $(pk_u,sk_u),(pk_v,sk_v)\leftarrow \textnormal{\textbf{KA.Gen}}(\textsf{KApp})$, we have $\textnormal{\textbf{KA.Agree}}(sk_u,pk_v)=\textnormal{\textbf{KA.Agree}}(sk_v,pk_u).$

  \item \textbf{Security in Honest-but-Curious Model.} For any $(pk_u,sk_u),(pk_v,sk_v)\leftarrow \textnormal{\textbf{KA.Gen}}(\textsf{KApp})$ and $s_{u,v}\leftarrow \textnormal{\textbf{KA.Agree}}(sk_u,pk_v)$, we have
      $$s_{u,v}\approx_c r$$
      in the view of any probabilistic polynomial-time (PPT) adversary who is given the public keys $(pk_u,pk_v)$,
      where $r$ is a uniformly random string and ``$\approx_c$'' represents that the two distributions are computationally indistinguishable.
\end{enumerate}
Like \cite{BIK17}, we will use the Diffie-Hellman key agreement scheme \cite{di76}, composed with a hash function.   That is,
$\textnormal{\textbf{KA.Setup}}(1^\lambda)\rightarrow (\mathbb{G},q,g,H)$ samples a group $\mathbb{G}$ of prime order $q$, along with a generator $g$, and a hash function $H$; $\textnormal{\textbf{KA.Gen}}(\mathbb{G},q,g,H)\rightarrow (x_u,g^{x_u})$ samples a random element $x_u\leftarrow \mathbb{Z}_q$ as the secret key $sk_u$ and computes $g^{x_u}$ as the public key $pk_u$; and $\textnormal{\textbf{KA.Agree}}(x_u,g^{x_v})\rightarrow s_{u,v}$ outputs $s_{u,v}=H((g^{x_v})^{x_u})$.

\subsection{Authenticated Encryption}
An authenticated encryption scheme contains polynomial-time algorithms $\textbf{AE} =(\textbf{AE.Setup}, \textbf{AE.Enc}, \textbf{AE.Dec})$, defined as follows:
\begin{itemize}
  \item $\textnormal{\textbf{AE.Setup}}(1^\lambda)\rightarrow sk$: outputs a secret key $sk$.
  \item $\textnormal{\textbf{AE.Enc}}(sk,m) \rightarrow c$: outputs a ciphertext $c$, where $m$ is a message.
  \item $\textnormal{\textbf{AE.Dec}}(sk,c)\rightarrow m\slash \bot$: outputs the message $m$ or $\bot$.
\end{itemize}
The above authenticated encryption scheme satisfies the following requirements:
\begin{enumerate}
  \item \textbf{Correctness.} For any $sk\leftarrow \textnormal{\textbf{AE.Setup}}(1^\lambda)$ and any message $m$, we have $\textnormal{\textbf{AE.Dec}}(sk,\textnormal{\textbf{AE.Enc}}(sk,m))=m.$

  \item \textbf{Security.} In general, we say an authenticated encryption is secure if it satisfies indistinguishability under chosen plaintext attacks (IND-CPA security) and ciphertext integrity.  Here, we omit the specific descriptions of the two security definitions, and refer the reader to \cite{BN00} for details.
\end{enumerate}

\section{Technical Overview}
\label{Sec:3}
Like previous secure aggregation protocols \cite{BIK17,CSH20,b20}, our protocol consists of a single server and $n$ clients $\mathcal{U}$, where we assume that each client $u\in \mathcal{U}$ has its own input vector $\mathbf{x}_u$, and the goal of the protocol is to compute $\mathbf{x}=\sum_{u \in \mathcal{U}}\mathbf{x}_u$ while making sure that the clients learn nothing more than their own input vectors and the server only learns the sum $\mathbf{x}$.

\subsection{Secure Aggregation with Shamir's Secret Sharing}
\label{Sec:3.1}
Our observation is that $\mathbf{x}=\sum_{u \in \mathcal{U}}\mathbf{x}_u$ can be computed using \textbf{SS} scheme.  We assume that $\mathbf{x}_u=(\mathbf{x}_{u,1}, \ldots, \mathbf{x}_{u,m})$.  For $i\in[m]$,  the client $u \in \mathcal{U}$ shares each element $\mathbf{x}_{u,i}$ with all other clients using \textbf{SS} scheme:  $\{[\![\mathbf{x}_{u,i}]\!]_v\}_{v\in \mathcal{U}\setminus\{u\}}\leftarrow \textbf{SS.Share}(t,\mathcal{U}\setminus\{u\},\mathbf{x}_{u,i})$.  After obtaining the shares $\{[\![\mathbf{x}_{v,i}]\!]_u\}_{v\in \mathcal{U}\setminus\{u\},i\in [m]}$ from other clients, the client $u \in \mathcal{U}$ adds up these shares: $[\![\mathbf{x}_i]\!]_u=\sum_{v \in \mathcal{U}}[\![\mathbf{x}_{v,i}]\!]_u$, and then sends $[\![\mathbf{x}_i]\!]_u$ to the server.  For $i\in[m]$, the server reconstructs $\mathbf{x}_i \leftarrow \textbf{SS.Recon}\big(t, \{[\![\mathbf{x}_i]\!]_u\}_{u\in\mathcal{U}}\big)$ such that $\mathbf{x}_i=\sum_{u\in \mathcal{U}}\mathbf{x}_{u,i}$ due to the linearity property of the \textbf{SS} scheme.  Let $\mathbf{x}=(\mathbf{x}_1,\ldots,\mathbf{x}_m)$, we have  $\mathbf{x}=\sum_{u \in \mathcal{U}}\mathbf{x}_u$.

Although the above approach successfully handles up to $n-t$ dropouts, the obvious drawback is that each client needs to share their own input element $m$ times and the server needs to reconstruct the shares $m$ times, i.e., running \textbf{SS.Share} and \textbf{SS.Recon} algorithms $m$ times, respectively.  Note that this approach is similar to the generic MPC protocol based on secret sharing.

\subsection{Ramp Secret Sharing to Improve Efficiency}
\label{Sec:3.2}
To address the above efficiency issue, we exploit an efficient $(t,d,n)$-ramp secret sharing \textbf{RSS} (cf. Section \ref{Sec:2.2}) to achieve secure aggregation.

Given the input vector $\mathbf{x}_u$, each client $u \in \mathcal{U}$ first divides $\mathbf{x}_u$ into $\lceil m/d\rceil$ vectors, i.e., $\mathbf{x}_u=(\mathbf{x}^1_{u},\ldots,\mathbf{x}^{\lceil m/d\rceil}_{u})$, and then for $i\in [\lceil m/d\rceil]$, the client $u \in \mathcal{U}$ shares each vector $\mathbf{x}^i_{u}$ with all other clients using \textbf{RSS} scheme.  The rest of the procedure is the same as in Section \ref{Sec:3.1}.  Consequently, each client and the server only need to call \textbf{RSS.Share} and \textbf{RSS.Recon} algorithms $\lceil m/d\rceil$ times, respectively.

The larger the size $d$ of the secret, the higher the computation and communication efficiency.  However, to achieve Shannon perfect security, we require that at most $t-d$ clients can be corrupted.  Therefore, there is a tradeoff between the size of the secret $d$, the threshold $t$, and the number of dropouts ($\leq n-t$).  Additionally, within the acceptable range of security (in many practical applications, Shannon perfect security is too much security), we can keep the parameter $d$ as close to $t$ as possible to improve efficiency.

\section{The FSSA Protocol}
\label{Sec:4}
We construct FSSA, a new secure aggregation protocol consisting of 3 rounds of communication between each client and the server.  We assume that each client $u \in \mathcal{U}$ has input vector $\mathbf{x}_u$ that is in $\mathbb{Z}^m_B$ for integers $m$ and $B$, and $\mathbb{Z}_q$ for some large prime $q$ is the finite field of the \textbf{RSS} scheme.   Clients cannot communicate with each other, but can communicate with the server via some secure channel (e.g., using cryptographic protocols such as TLS).  In FSSA, clients may drop out of the protocol in any round and the server can compute a correct aggregated vector as long as the number of dropouts is at most $n-t$ (i.e., at least $t$ clients survive to the last round).  To facilitate the use of the \textbf{RSS} scheme, we also assume that each client is assigned a unique index $u\in [n]$.  We provide the complete description of FSSA in Figure \ref{f1}.


\begin{figure*}
\begin{center}
FSSA Protocol
\end{center}
\begin{itemize}
  \item \textbf{Setup.} All clients are given public parameters $pp:=(\lambda, n, t, d, \mathbb{Z}_q, \mathbb{Z}^m_B, \textsf{KApp})$, where $n$ is the number of clients (let $\mathcal{U}$ be the set of all clients and $|\mathcal{U}|=n$), $t$, $d$ and $\mathbb{Z}_q$ for some large prime $q$ are the threshold value, the size of the secret and finite field of the \textbf{RSS} scheme respectively, $\mathbb{Z}^m_B$ is the input domain, and $\textsf{KApp} \leftarrow \textnormal{\textbf{KA.Setup}}(1^\lambda)$.
      \vspace{0.2cm}
  \item \textbf{Round 0 (AdvertiseKeys)}.\\
Client $u$:
\begin{itemize}
  \item Generate $(sk_u, pk_u)\leftarrow \textnormal{\textbf{KA.Gen}}(\textsf{KApp})$.
  \item Send $pk_u$ to the server.
\end{itemize}
Server:
\begin{itemize}
  \item Collect public keys from at least $t$ clients (let $\mathcal{U}_1\subseteq \mathcal{U}$ be the set of these clients).
  \item Broadcast $\{(u,pk_u)\}_{u\in \mathcal{U}_1}$ to all clients in $\mathcal{U}_1$.
\end{itemize}
\vspace{0.2cm}
  \item \textbf{Round 1 (ShareVector)}:\\
  Client $u$:
  \begin{itemize}
    \item After receiving $\{(u,pk_u)\}_{u\in \mathcal{U}_1}$, check whether $|\mathcal{U}_1|\geq t$ and all public keys are different.  Otherwise, abort.
    \item Divide the input vector $\mathbf{x}_u\in \mathbb{Z}^m_B$ into $\lceil m/d\rceil$ vectors, i.e., $\mathbf{x}_u=(\mathbf{x}^1_{u},\ldots,\mathbf{x}^{\lceil m/d\rceil}_{u})$, where $\mathbf{x}^i_{u}\in\mathbb{Z}^d_R$ for $i\in[\lceil m/d\rceil-1]$ and $\mathbf{x}^{\lceil m/d\rceil}_{u}\in \mathbb{Z}^{\leq d}_R$.
    \item For each $i\in[\lceil m/d\rceil]$, generate shares of $\mathbf{x}^i_{u}$:  $\{[\![\mathbf{x}^i_{u}]\!]_v\}_{v\in \mathcal{U}_1}\leftarrow \textbf{RSS.Share}(t,d,\mathcal{U}_1,\mathbf{x}^i_{u})$.
     \item For each other client $v\in \mathcal{U}_1\setminus\{u\}$ and $i\in[\lceil m/d\rceil]$, generate the pairwise symmetric key $s_{u,v}\leftarrow \textbf{KA.Agree}(sk_u,pk_v)$, and the ciphertext $C_{u\rightarrow v}\langle i\rangle\leftarrow \textbf{AE.Enc}\big(s_{u,v}, (u,v,[\![\mathbf{x}^i_{u}]\!]_v)\big)$.
      \item Send all ciphertexts $\{C_{u \rightarrow v}\langle i\rangle\}_{v\in \mathcal{U}_1\setminus\{u\}, i\in[\lceil m/d\rceil]}$ to the server.  In addition, store all messages received and values generated in this round.
  \end{itemize}
  Server:
\begin{itemize}
  \item Collect ciphertexts from at least $t$ clients (let $\mathcal{U}_2\subseteq \mathcal{U}_1$ be the set of these clients).
  \item Send $\{(v,C_{v\rightarrow u}\langle i\rangle)\}_{v\in \mathcal{U}_2\setminus\{u\}, i\in[\lceil m/d\rceil]}$ to each client $u \in \mathcal{U}_2$.
\end{itemize}
      \vspace{0.2cm}
  \item \textbf{Round 2 (Reconstruction and Aggregation)}: \\
  Client $u$:
  \begin{itemize}
    \item After receiving $\{(v,C_{v\rightarrow u}\langle i\rangle)\}_{v\in \mathcal{U}_2\setminus\{u\}, i\in[\lceil m/d\rceil]}$, check whether $|\mathcal{U}_2|\geq t$. Otherwise, abort.
    \item For each other client $v\in \mathcal{U}_2\setminus\{u\}$, decrypt all ciphertexts: $(u',v', [\![\mathbf{x}^i_{v}]\!]_u)\leftarrow \textbf{AE.Dec}\big(\textbf{KA.Agree}(sk_u,pk_v), C_{v\rightarrow u}\langle i\rangle\big)$ and check whether $u=u'\wedge v=v'$ for $i\in[\lceil m/d\rceil]$.  Otherwise, abort.
    \item For $i\in[\lceil m/d\rceil]$, compute $[\![\mathbf{x}^i]\!]_u=\sum_{v \in \mathcal{U}_2}[\![\mathbf{x}^i_{v}]\!]_u$.  Send $\{[\![\mathbf{x}^i]\!]_u\}_{i\in[\lceil m/d\rceil]}$ to the server.
  \end{itemize}
  Server:
  \begin{itemize}
    \item Collect messages from at least $t$ clients (let $\mathcal{U}_3\subseteq \mathcal{U}_2$ be the set of these clients).
    \item For each $i\in[\lceil m/d\rceil]$, reconstruct $\mathbf{x}^i \leftarrow \textbf{RSS.Recon}\big(t, d, \{[\![\mathbf{x}^i]\!]_u\}_{u\in\mathcal{U}_3}\big)$ such that $\mathbf{x}^i=\sum_{u\in \mathcal{U}_2}\mathbf{x}^i_u$ due to the linearity property of the \textbf{RSS} scheme.
    \item Output the aggregated vector $\mathbf{x}=(\mathbf{x}^1,\ldots, \mathbf{x}^{\lceil m/d\rceil})$.
  \end{itemize}

\end{itemize}
\centering
\caption{Detailed description of the FSSA protocol}
\label{f1}
\end{figure*}

\section{Correctness and Security Analysis}
\label{Sec:5}
In this section, we analyze the correctness and security of FSSA.
\subsection{Correctness of FSSA}
\label{Sec:5.1}
The correctness of FSSA requires that the server should output a correct aggregated vector at the end of the protocol as long as each client and the server run the protocol honestly and at least $t$ clients survive to the last round.
\begin{theorem}[Correctness] \label{Te1} Given $pp:=(\lambda, n, t, d, \mathbb{Z}_q, \mathbb{Z}^m_B, \textsf{KApp})$, where $B< q$.  Let $R=n(B-1)+1$ satisfying $R\leq q$.  For input vectors $\{\mathbf{x}_u\in \mathbb{Z}^m_B\}_{u\in \mathcal{U}}$, where $\mathbf{x}_u=(\mathbf{x}^1_{u},\ldots,\mathbf{x}^{\lceil m/d\rceil}_{u})$,  if $|\mathcal{U}_3|\geq t$, then the server can compute and output the aggregated vector $\mathbf{x}=(\mathbf{x}^1,\ldots, \mathbf{x}^{\lceil m/d\rceil})$, where $\mathbf{x}^i=\sum_{u\in \mathcal{U}_2}\mathbf{x}^i_u$.
\end{theorem}
\begin{proof}
By the correctness of the \textbf{KA} and \textbf{AE} schemes, each client $u\in\mathcal{U}_2$ can obtain the shares $\{[\![\mathbf{x}^i_{v}]\!]_u\}_{v\in \mathcal{U}_2}$, where $\{[\![\mathbf{x}^i_{v}]\!]_u\}_{u\in \mathcal{U}_2}\leftarrow \textbf{RSS.Share}(t,d,\mathcal{U}_2,\mathbf{x}^i_{v})$ for $i\in[\lceil m/d\rceil]$.  Then, after receiving the vectors $\{[\![\mathbf{x}^i]\!]_u=\sum_{v \in \mathcal{U}_2}[\![\mathbf{x}^i_{v}]\!]_u\}_{u\in\mathcal{U}_3, i\in[\lceil m/d\rceil]}$, the server obtains the vector $\mathbf{x}^i$ by running $\mathbf{x}^i \leftarrow \textbf{RSS.Recon}\big(t, d, \{[\![\mathbf{x}^i]\!]_u\}_{u\in\mathcal{U}_3}\big)$ for $i\in[\lceil m/d\rceil]$.  By the linearity property of the \textbf{RSS} scheme, we have $$\mathbf{x}^i=\sum_{u\in \mathcal{U}_2}\mathbf{x}^i_u$$ for $i\in[\lceil m/d\rceil]$, which completes the proof.

\end{proof}
\subsection{Threat Model of FSSA}
\label{Sec:5.2}
In FSSA, we want to prevent each client's input vector (i.e., gradient in FL) from being leaked to other clients and the server, while ensuring that the server obtains a correct aggregated vector at the end of the protocol.  As with the previous secure aggregation protocols \cite{BIK17,so21}, to preserve the privacy of the input vectors, we consider honest-but-curious setting (a.k.a. semi-honest setting), where both the clients and the server follow the protocol honestly, but attempt to infer information about input vectors of the other clients.

Since both the clients and the server are honest-but-curious, we consider two security notions: one about honest-but-curious clients, where any set of less than $t-d$ clients collude (excluding the server) with the aim to infer information about input vectors of the other clients; the other is about honest-but-curious server, where the server colludes with any set of less than $t-d$ clients with the purpose of inferring information about input vectors of the other clients.

For completeness, we also give the following description for active adversaries.  By active adversaries, we mean parties (clients or the server) that deviate from the protocol, sending incorrect and\slash or arbitrarily
chosen messages to honest clients, omitting messages, aborting, and sharing their entire view of the protocol with each other, and also
with the server (if the server is also an active adversary).   Like \cite{BIK17}, we can only show input privacy for honest clients, as it is much harder to additionally guarantee correctness and availability for the protocol when some clients are actively adversarial, and we also require the support of a public key infrastructure that allows clients to register identities, and sign messages using their identity, such that other clients can verify this signature, but cannot impersonate them.   We omit the details of how to achieve the privacy against active adversaries because the construction idea and technique used are the same as \cite{BIK17}.
\subsection{Security of FSSA}
\label{Sec:5.3}
As stated in the introduction, secure aggregation is essentially a multi-party computation.  Therefore, we use a simulation-based proof which is commonly used in secure MPC protocols to prove that neither the clients nor the server could learn any information about inputs of the other clients.

For honest-but-curious clients, we show that the joint view of any subset of $\leq t-d$ clients can be simulated given only the inputs of these clients.  This indicates that these honest-but-curious clients learn ``nothing more'' than their own inputs.

 For honest-but-curious server, in addition to the aggregated vector, the server has information about inputs of some honest-but-curious clients ($\leq t-d$ clients), so we show that the joint view of such collusion can be simulated given only the inputs of these honest-but-curious clients and the sum of inputs of the remaining clients.   This indicates that the server and these honest-but-curious clients learn ``nothing more'' than the sum of inputs of the other clients and their own inputs.

For simplicity, let $\mathbf{x}_{\mathcal{U}}=\{\mathbf{x}_u\}_{u\in\mathcal{U}}$ denotes the collection of inputs from the clients in $\mathcal{U}$.  Given any subset $\mathcal{C}\subseteq \mathcal{U}\cup \{\text{Server}\}$, let $\textnormal{\textbf{REAL}}^{\mathcal{U},\lambda,t,d}_{\mathcal{C}}(\mathbf{x}_{\mathcal{U}}, \mathcal{U}_1,\mathcal{U}_2,\mathcal{U}_3)$ be a random variable representing the joint view of all parties in $\mathcal{C}$, where the randomness is over the internal
randomness of all parties, and the randomness in the setup phase.

 \begin{theorem}[Security Against Honest-but-Curious Clients]\label{Te2} Under the same parameter selections as Theorem \ref{Te1}, there exists a PPT simulator $\textnormal{\textbf{SIM}}$ such that for all clients $\mathcal{U}$ with inputs $\mathbf{x}_{\mathcal{U}}$ and $|\mathcal{U}|\geq t$, sets of clients $\mathcal{U}_1$, $\mathcal{U}_2$, $\mathcal{U}_3$, and $\mathcal{C}$ with inputs $\mathbf{x}_{\mathcal{C}}$ such that $\mathcal{U}_3\subseteq\mathcal{U}_2\subseteq\mathcal{U}_1\subseteq\mathcal{U}$, $\mathcal{C}\subseteq \mathcal{U}$ and $|\mathcal{C}|\leq t-d$, the output of
$\textnormal{\textbf{SIM}}^{\mathcal{U},\lambda,t,d}_{\mathcal{C}}$ is computationally indistinguishable from the output of $\textnormal{\textbf{REAL}}^{\mathcal{U},\lambda,t,d}_{\mathcal{C}}:$
$$\textnormal{\textbf{SIM}}^{\mathcal{U},\lambda,t,d}_{\mathcal{C}}(\mathbf{x}_{\mathcal{C}}, \mathcal{U}_1,\mathcal{U}_2,\mathcal{U}_3) \approx_c \textnormal{\textbf{REAL}}^{\mathcal{U},\lambda,t,d}_{\mathcal{C}}(\mathbf{x}_{\mathcal{U}}, \mathcal{U}_1,\mathcal{U}_2,\mathcal{U}_3).$$

\end{theorem}
\begin{proof}
Since in this case the server is not involved, the joint view of the clients in $\mathcal{C}$ does not depend on the inputs of the clients not in $\mathcal{C}$.  We prove this theorem via a series of consecutive hybrids, where the first hybrid is the real execution and the last one is the simulated execution.\vspace{0.2cm}

\noindent \textbf{$\bullet$ Hybrid 0.}~ This is the real execution $\textnormal{\textbf{REAL}}^{\mathcal{U},\lambda,t,d}_{\mathcal{C}}$. \vspace{0.2cm}

\noindent \textbf{$\bullet$ Hybrid 1.}~ We now change how the symmetric keys of the honest clients in $\mathcal{U}_2\setminus \mathcal{C}$ are generated.  Specifically, instead of generating the symmetric keys $s_{u,v}$ by running $s_{u,v}\leftarrow \textbf{KA.Agree}(sk_u,pk_v)$ for $v\in \mathcal{U}_2\setminus \mathcal{C}$, we choose uniformly random values $s'_{u,v}$ as the keys.  By the security of the $\textbf{KA}$ scheme, we conclude that Hybrid 1 is computationally indistinguishable from Hybrid 0. \vspace{0.2cm}

\noindent \textbf{$\bullet$ Hybrid 2.}~ We now replace all ciphertexts encrypted by the honest clients in $\mathcal{U}_2\setminus \mathcal{C}$ sent to other honest clients with encryptions of $0$ instead of shares of their inputs.  By IND-CPA security of the $\textbf{AE}$ scheme,  we conclude that Hybrid 2 is computationally indistinguishable from Hybrid 1. \vspace{0.2cm}

\noindent \textbf{$\bullet$ Hybrid 3.}~We now replace the shares of inputs of honest clients in $\mathcal{U}_2\setminus \mathcal{C}$ sent to the corrupted clients in $\mathcal{C}$ with shares of 0.  The Shannon perfect security of the \textbf{RSS} scheme (due to $|\mathcal{C}|\leq t-d$) ensures that the distribution of any $|\mathcal{C}|$ shares of any vector is identical to that of $|\mathcal{C}|$ shares of 0.  It
follows that Hybrid 3 is identically distributed to Hybrid 2.

We now define a PPT simulator \textnormal{\textbf{SIM}} that samples from the distribution described in Hybrid 3.  By transitivity of indistinguishability, we conclude that the output of $\textnormal{\textbf{REAL}}^{\mathcal{U},\lambda,t,d}_{\mathcal{C}}$ is computationally indistinguishable from the output of $\textnormal{\textbf{SIM}}^{\mathcal{U},\lambda,t,d}_{\mathcal{C}}$.

\end{proof}

\begin{theorem}[Security Against Honest-but-Curious Server]\label{Te3} Under the same parameter selections as Theorem \ref{Te1}, there exists a PPT simulator $\textnormal{\textbf{SIM}}$ such that for all clients $\mathcal{U}$ with inputs $\mathbf{x}_{\mathcal{U}}$, sets of clients $\mathcal{U}_1$, $\mathcal{U}_2$, $\mathcal{U}_3$, and $\mathcal{C}$ with inputs $\mathbf{x}_{\mathcal{C}}$ such that $\mathcal{U}_3\subseteq\mathcal{U}_2\subseteq\mathcal{U}_1\subseteq\mathcal{U}$, $\mathcal{C}\subseteq \mathcal{U}\cup\{\text{Server}\}$ and $|\mathcal{C}\setminus \{\text{Server}\}|\leq t-d$, the output of
$\textnormal{\textbf{SIM}}^{\mathcal{U},\lambda,t,d}_{\mathcal{C}}$ is computationally indistinguishable from the output of $\textnormal{\textbf{REAL}}^{\mathcal{U},\lambda,t,d}_{\mathcal{C}}:$
$$\textnormal{\textbf{SIM}}^{\mathcal{U},\lambda,t,d}_{\mathcal{C}}(\mathbf{x}_{\mathcal{C}}, \mathbf{y}, \mathcal{U}_1,\mathcal{U}_2,\mathcal{U}_3) \approx_c \textnormal{\textbf{REAL}}^{\mathcal{U},\lambda,t,d}_{\mathcal{C}}(\mathbf{x}_{\mathcal{U}}, \mathcal{U}_1,\mathcal{U}_2,\mathcal{U}_3),$$
where  \begin{equation*}\label{eq1}
        \mathbf{y}=\left\{
 \begin{array}{lllll}
  \sum_{u\in\mathcal{U}_2\setminus \mathcal{C}}\mathbf{x}_{u}, & \text{if}~ |\mathcal{U}_2|\geq t,\\
   \bot, & \text{otherwise}.
  \end{array}\right.
  \end{equation*}
\end{theorem}
\begin{proof}
The proof of this theorem is similar to that of Theorem \ref{Te2}, with the main difference being that the sum of inputs of the clients in $\mathcal{U}_2\setminus \mathcal{C}$ affects the distribution of inputs of the honest clients.\\

\noindent \textbf{$\bullet$ Hybrid 0.}~ This is the real execution $\textnormal{\textbf{REAL}}^{\mathcal{U},\lambda,t,d}_{\mathcal{C}}$. \vspace{0.2cm}

\noindent \textbf{$\bullet$ Hybrid 1.}~ We now change how the symmetric keys of the honest clients in $\mathcal{U}_2\setminus \mathcal{C}$ are generated.  Specifically, instead of generating the symmetric keys $s_{u,v}$ by running $s_{u,v}\leftarrow \textbf{KA.Agree}(sk_u,pk_v)$ for $v\in \mathcal{U}_2\setminus \mathcal{C}$, we choose uniformly random values $s'_{u,v}$ as the keys.  By the security of the $\textbf{KA}$ scheme, we have that Hybrid 1 is computationally indistinguishable from Hybrid 0. \vspace{0.2cm}

\noindent \textbf{$\bullet$ Hybrid 2.}~ We now replace all ciphertexts encrypted by the honest clients in $\mathcal{U}_2\setminus \mathcal{C}$ sent to other honest clients with encryptions of $0$ instead of shares of their inputs.  By the IND-CPA security of the $\textbf{AE}$ scheme,  we conclude that Hybrid 2 is computationally indistinguishable from Hybrid 1. \vspace{0.2cm}

\noindent \textbf{$\bullet$ Hybrid 3.}~In this hybrid, for each honest client $u\in \mathcal{U}_2\setminus \mathcal{C}$ with input $\mathbf{x}_u=(\mathbf{x}^1_{u},\ldots,\mathbf{x}^{\lceil m/d\rceil}_{u})$, we substitute the shares of $\mathbf{x}^i_{u}$ sent to the corrupted clients in $\mathcal{C}$ (in \textbf{Round 1} of FSSA) with shares of $\mathbf{z}^i_{u}$ for $i\in[\lceil m/d\rceil]$, which are sampled depending on $\mathbf{y}$, as follows:
\begin{enumerate}
  \item If $\mathbf{y}=\bot$, then we let $\mathbf{z}^i_{u}=\mathbf{0}$ for $i\in[\lceil m/d\rceil]$.  In this case, $\mathbf{y}=\bot$ means that $|\mathcal{U}_2|< t$, and hence $|\mathcal{U}_3|< t$.
  \item  Otherwise, we choose uniformly random vectors $\{\mathbf{z}^i_{u}\}_{\mathcal{U}_2\setminus \mathcal{C},i\in[\lceil m/d\rceil]}$ subject to $\sum_{u\in \mathcal{U}_2\setminus \mathcal{C}}\mathbf{z}_{u}=\mathbf{y}$, where $\mathbf{z}_{u}=(\mathbf{z}^1_{u},\ldots,\mathbf{z}^{\lceil m/d\rceil}_{u})$.
\end{enumerate}

In summary, the joint view of the corrupted parties in $\mathcal{C}$ contains only $|\mathcal{C}|\leq t-d$ shares.  By the Shannon perfect security of the \textbf{RSS} scheme (due to $|\mathcal{C}|\leq t-d$), we have that the distribution of any $|\mathcal{C}|$ shares of the vector $\mathbf{x}^i_{u}$ is identical to that of $|\mathcal{C}|$ shares of the vector $\mathbf{z}^i_{u}$, thus making Hybrid 3 identically distributed to Hybrid 2.

%
%
%
%

We now define a PPT simulator \textnormal{\textbf{SIM}} that samples from the distribution described in Hybrid 3.  By transitivity of indistinguishability, we conclude that the output of $\textnormal{\textbf{REAL}}^{\mathcal{U},\lambda,t,d}_{\mathcal{C}}$ is computationally indistinguishable from the output of $\textnormal{\textbf{SIM}}^{\mathcal{U}, \lambda,t,d}_{\mathcal{C}}$.

\end{proof}
Like \cite{BIK17}, we can use a secure signature scheme to convert our FSSA protocol that is secure against honest-but-curious adversaries into one that is secure against active adversaries, but this requires an extra round of communication and a trusted third party to generate secret\slash public keys.  In this paper, we are only concerned with security in the honest-but-curious setting.

\section{Performance Evaluation}
\label{Sec:6}
In this section, we evaluate the performance of FSSA protocol.   Similar to \cite{BIK17}, in our experiments we use the following cryptographic primitives:
\begin{itemize}
  \item Shamir's $(t,n)$-secret sharing scheme to construct the \textbf{RSS} scheme; see below.
  \item Elliptic-Curve Diffie-Hellman over the NIST P-256 curve with SHA-256 to construct the \textbf{KA} scheme.
  \item AES-GCM with 256-bit keys to construct the \textbf{AE} scheme.
\end{itemize}

In \cite{Ta13}, the authors presented the following simple construction of the \textbf{RSS} scheme using Shamir's $(t,n)$-secret sharing scheme:  given a secret $\mathbf{s}=(s_0,s_1,\ldots,s_{d-1})$, choose $t-d$ uniformly random $a_i\in \mathcal{M}$ and set $f(x)=s_0+s_1x+\cdots+s_{d-1}x^{d-1}+a_dx^d+\cdots+a_{t-1}x^{t-1}$.  Then, with the polynomial $f(x)$, we can use Shamir's $(t,n)$-secret sharing to generate $n$ shares and reconstruct the secret $\mathbf{s}$ from at least $t$ shares.   In this work, we use the above construction of the $(t,d,n)$-ramp secret sharing scheme in the experimental evaluation of our protocol.

We assume that each element of each client's data vector is an integer that requires at most 16-bit of storage, i.e., $B=2^{16}$; prior work \cite{GAGN15} shows that deep networks can be trained using only $16$-bit wide fixed-point number representation to achieve near-lossless accuracy.   For $B=2^{16}$, we can set $q=2^{25}$ for any $n\leq 500$ by Theorem \ref{Te1}.  This means that we can encrypt the concatenated elements $[\![\mathbf{x}^1_{u}]\!]_v||\cdots||[\![\mathbf{x}^{\lceil 128/25 \rfloor}_{u}]\!]_v$ using the \textbf{AE} scheme in \textbf{Round 1} of FSSA, i.e.,
$C_{u\rightarrow v}\langle i\rangle\leftarrow \textbf{AE.Enc}\big(s_{u,v}, (u,v,[\![\mathbf{x}^1_{u}]\!]_v||\cdots||[\![\mathbf{x}^5_{u}]\!]_v)\big)$.

In addition, note that the computation cost of the \textbf{RSS} scheme consists of 1) generating $n$ shares of the secret, which
is the same as that of Shamir's $(t,n)$-secret sharing, and 2) reconstructing the secret, which requires $O(n)$ computation by precomputing the Lagrange basis polynomials:
$$L_{u}(x)=\prod_{v\in \mathcal{U}\setminus\{u\}}\frac{(x-v)}{(u-v)}~(\textnormal{mod}~q)$$ for any client $u\in \mathcal{U}$.

We run single-threaded simulations on a virtual Linux environment based on a workstation with an Intel(R) Core(TM) i5-8250U CPU @1.60GHz and 16.0GB RAM, and implement FSSA in Python.  Since our FSSA takes into account dropouts and corruptions, our experimental results will be shown around these two variables.  For simplicity, we assume that clients may drop out of the protocol after sending the public keys to the server but before sending the shares of their data vectors to all other clients, i.e., we have $|\mathcal{U}_1|=n$ and $|\mathcal{U}_2|\leq n$, and we let $\rho$ and $\gamma$ denote client dropout rate and client corruption rate, respectively.  In addition, in our experiments the threshold parameter $t$ will be set to $t=n-\rho n$.   To achieve Shannon perfect security of the \textbf{RSS} scheme, given the number of clients $n$, the client corruption rate $\gamma$ and the client dropout rate $\rho$, we compute the parameter $d=n-\gamma n-\rho n$ (recall that Shannon perfect security requires $\gamma n\leq t-d$).  For example, given $n=100$, $\rho=30\%$ and $\gamma=30\%$, we have $d=40$.  On the other hand, if we consider $(t-d)$-Privacy that is weaker than Shannon perfect security, we can obtain bigger parameter $d$,  thus producing better experimental results.  This is a tradeoff between efficiency and security.  Note that for the fixed $n$ and $\rho$, the higher $\gamma$, the smaller $d$.

In the following, we evaluate the computation and communication costs of FSSA under the following 4 cases:
\begin{itemize}
    \item \textbf{Case 1}: Different number of clients and client corruption rates with fixed data vector size and client dropout rate.
    \item \textbf{Case 2}: Different data vector sizes and client corruption rates with fixed number of clients and client dropout rate.
    \item \textbf{Case 3}: Different number of clients and client dropout rates with fixed data vector size and client corruption rate.
    \item \textbf{Case 4}: Different data vector sizes and client dropout rates with fixed number of clients and client corruption rate.
\end{itemize}

In terms of communication cost,  we only give the communication cost per client, and omit the communication cost of the server because it is essentially $n$ times the communication cost of each client.   We now discuss the above cases and provide detailed experiment results. In the following figures, each plotted point represents the average over 5 iterations.

 \begin{remark}\label{running}
Note that in Fig. \ref{fig1-4}, the curves are not stable, as the running times per client are not long enough to avoid being affected by other unrelated processes running during the experiment.  While in Fig. \ref{fig5-8}, the curves are extremely unstable, as the running times of the server are relatively small (millisecond scale) and therefore vulnerable to other unrelated processes running during the experiment.
\end{remark}


\subsection{Performance Analysis under Case 1}
\noindent \textbf{Computation Overhead.} Fig. \ref{fig1-4} and Fig. \ref{fig5-8} depict the running time of each client and the server,  respectively.  In general, it is sufficient to consider a maximum client dropout rate $\rho$ of $30\%$.  For the given $\rho$ of $30\%$, it is also sufficient to consider a maximum client corruption rate $\gamma$ of $30\%$, as such $\gamma$ is close to half of the proportion of undropped clients ($70\%$).

\renewcommand\thefigure{\arabic{figure}}
\setcounter{figure}{1}
\begin{figure}
\centering
\subfigure[When $\rho=30\%$] {
 \label{fig:1}
\includegraphics[width=0.45\columnwidth]{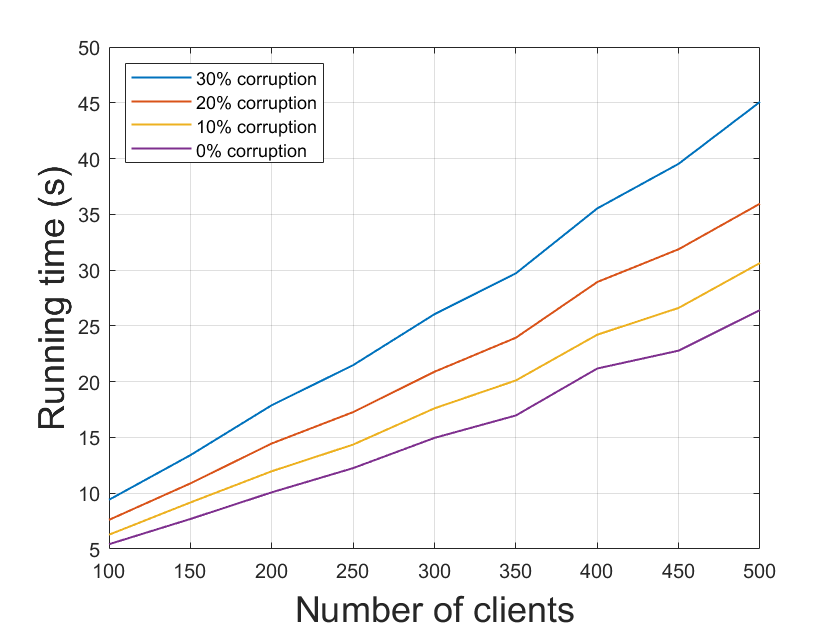}
}
\hspace{-0.2in}
\subfigure[When $\rho=20\%$] {
\label{fig:2}
\includegraphics[width=0.45\columnwidth]{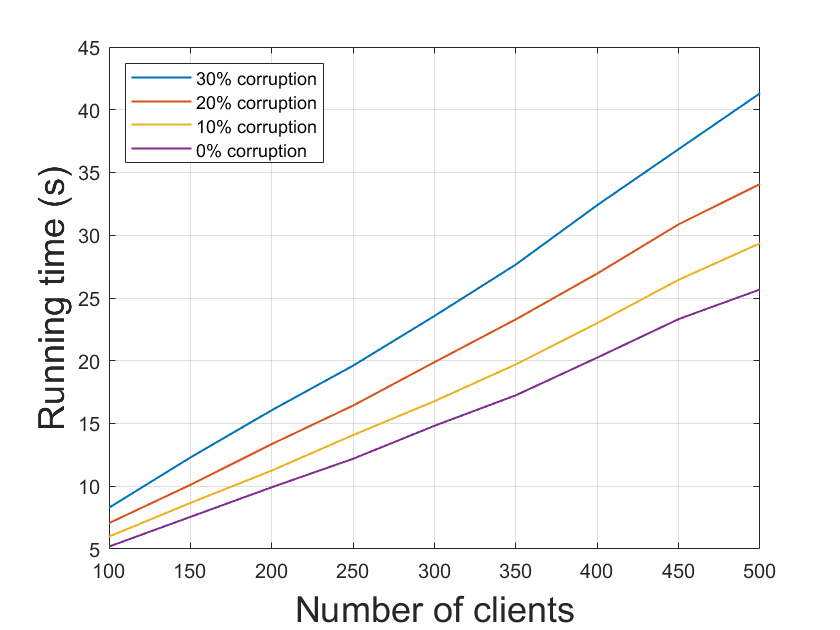}
}
\hspace{-0.2in}
\subfigure[When $\rho=10\%$] {
\label{fig:3}
\includegraphics[width=0.45\columnwidth]{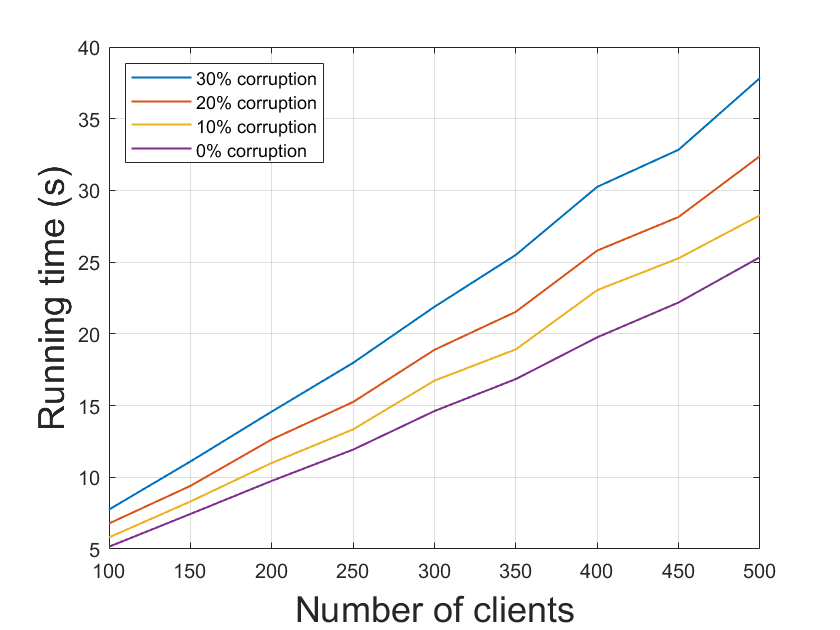}
}
\hspace{-0.2in}
\subfigure[When $\rho=0\%$] {
\label{fig:4}
\includegraphics[width=0.45\columnwidth]{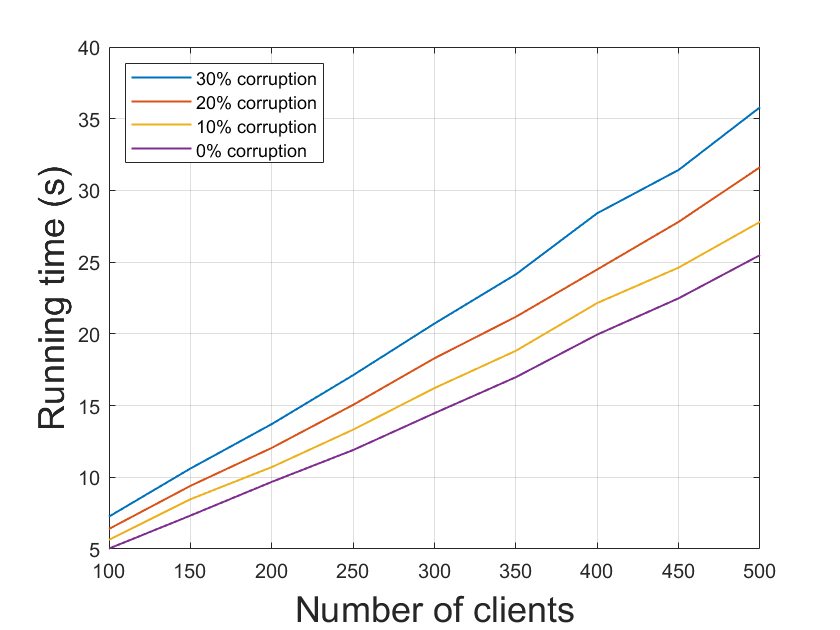}
}
\caption{Running time per client, where the data vector size is fixed to 100K.}
\label{fig1-4}
\end{figure}

\begin{figure}
\centering
\subfigure[When $\rho=30\%$] {
 \label{fig:5}
\includegraphics[width=0.45\columnwidth]{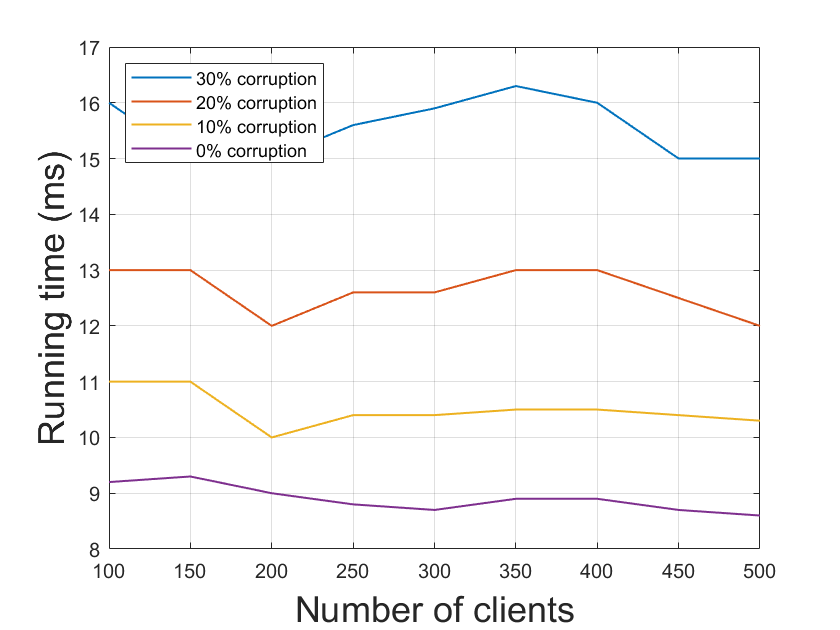}
}
\hspace{-0.2in}
\subfigure[When $\rho=20\%$] {
\label{fig:6}
\includegraphics[width=0.45\columnwidth]{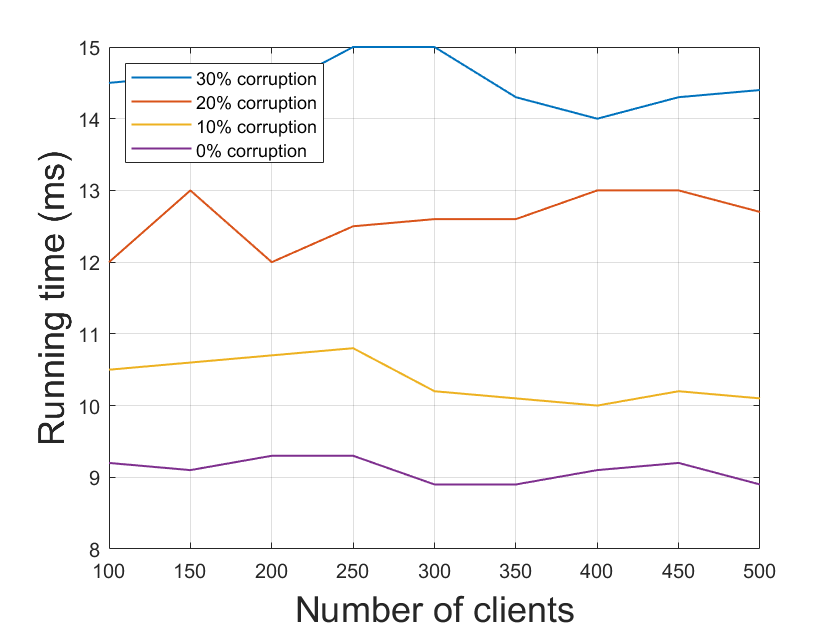}
}
\hspace{-0.2in}
\subfigure[When $\rho=10\%$] {
\label{fig:7}
\includegraphics[width=0.45\columnwidth]{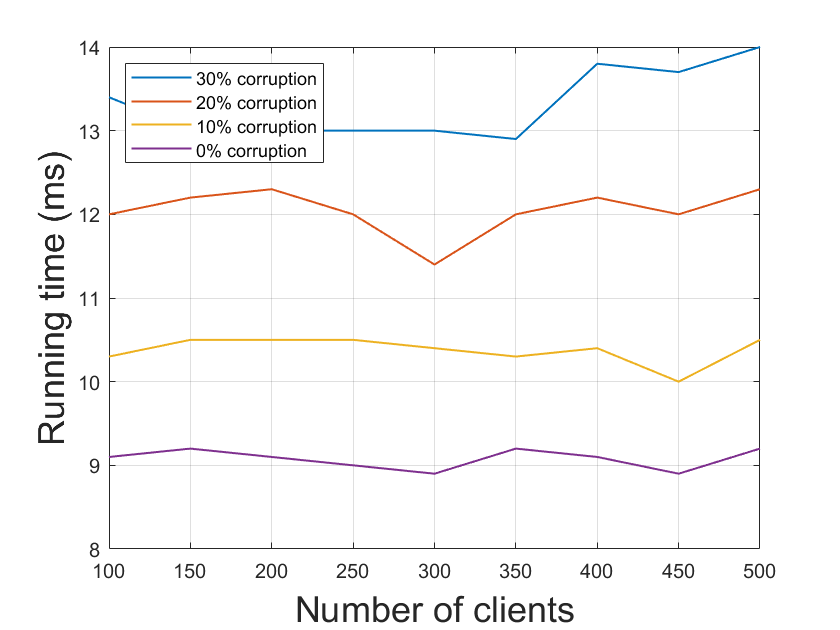}
}
\hspace{-0.2in}
\subfigure[When $\rho=0\%$] {
\label{fig:8}
\includegraphics[width=0.45\columnwidth]{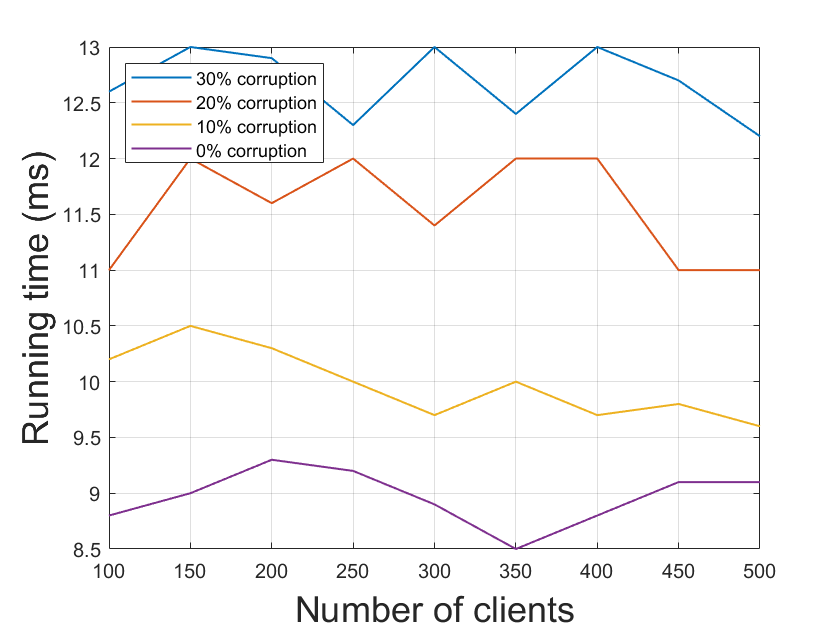}
}
\caption{Running time of the server, where the data vector size is fixed to 100K.}
\label{fig5-8}
\end{figure}

As seen in Fig. \ref{fig1-4} and Fig. \ref{fig5-8}, for the fixed $\rho$ and the data vector size, the running time of each client increases (approximately) linearly with $n$ and increases significantly with $\gamma$ for different $\rho$, while the running time of the server remains almost unchanged with $n$ but increases significantly with $\gamma$ for different $\rho$.  This is because the running time required for secret sharing (i.e., \textbf{RSS.Share} algorithm) is relatively long and is related to the parameter $d$, which is determined by $\gamma$ for the given $n$ and $\rho$; in comparison, the running time required for the secret reconstruction (i.e., \textbf{RSS.Recon} algorithm) is so short that $n$ has little impact on the computation cost of the server. \vspace{0.2cm}

\begin{figure}
\centering
\subfigure[When $\rho=30\%$] {
 \label{fig:-1}
\includegraphics[width=0.45\columnwidth]{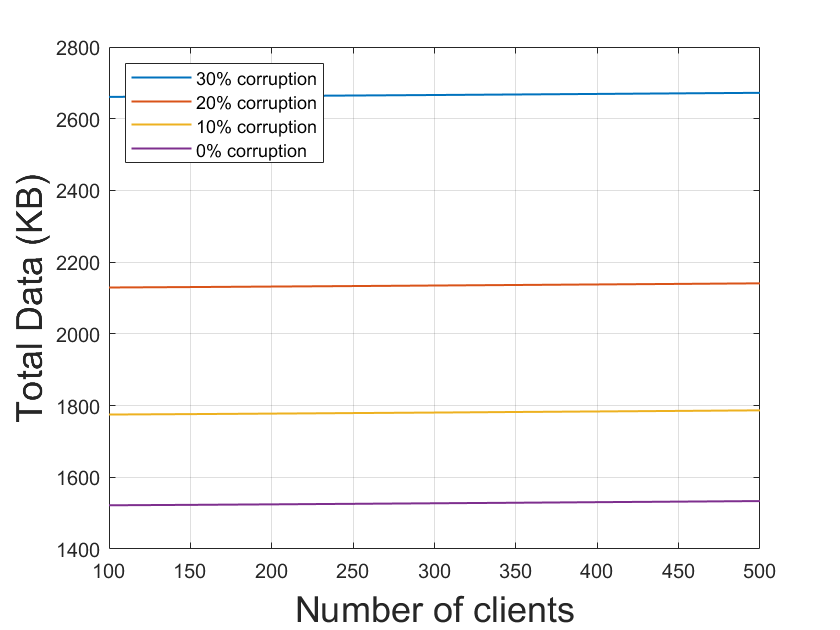}
}
\hspace{-0.2in}
\subfigure[When $\rho=20\%$] {
\label{fig:-2}
\includegraphics[width=0.45\columnwidth]{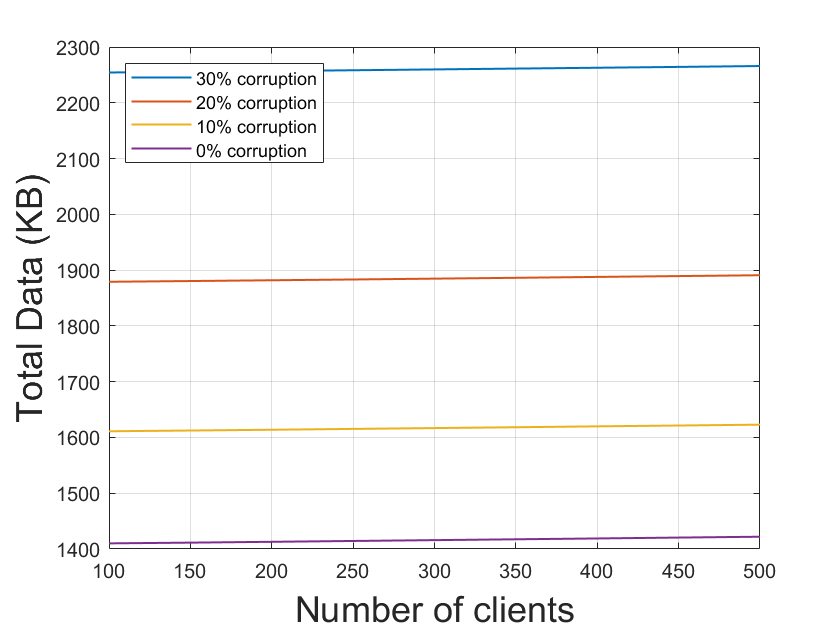}
}
\hspace{-0.2in}
\subfigure[When $\rho=10\%$] {
\label{fig:-3}
\includegraphics[width=0.45\columnwidth]{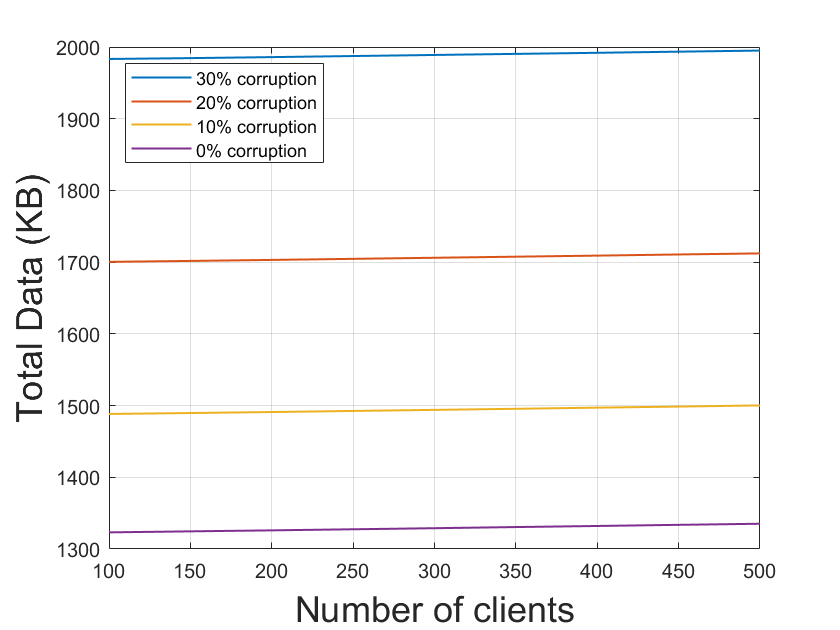}
}
\hspace{-0.2in}
\subfigure[When $\rho=0\%$] {
\label{fig:-4}
\includegraphics[width=0.45\columnwidth]{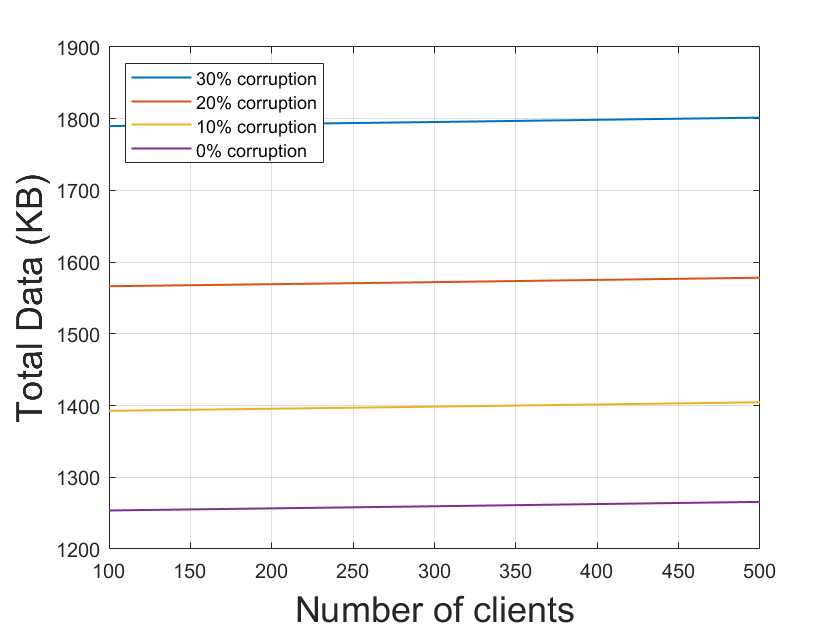}
}
\caption{Total data transfer per client, where the data vector size is fixed to 100K.}
\label{fig-1--4}
\end{figure}

\noindent \textbf{Communication Overhead.}  Fig. \ref{fig-1--4} shows the total amount of data transferred by each client.  We can see that, for the fixed $\rho$ and the data vector size, the communication cost of each client remains almost unchanged with $n$ but increases significantly with $\gamma$.

\subsection{Performance Analysis under Case 2}
\begin{figure}
\centering
\subfigure[When $\rho=30\%$] {
 \label{fig:9}
\includegraphics[width=0.45\columnwidth]{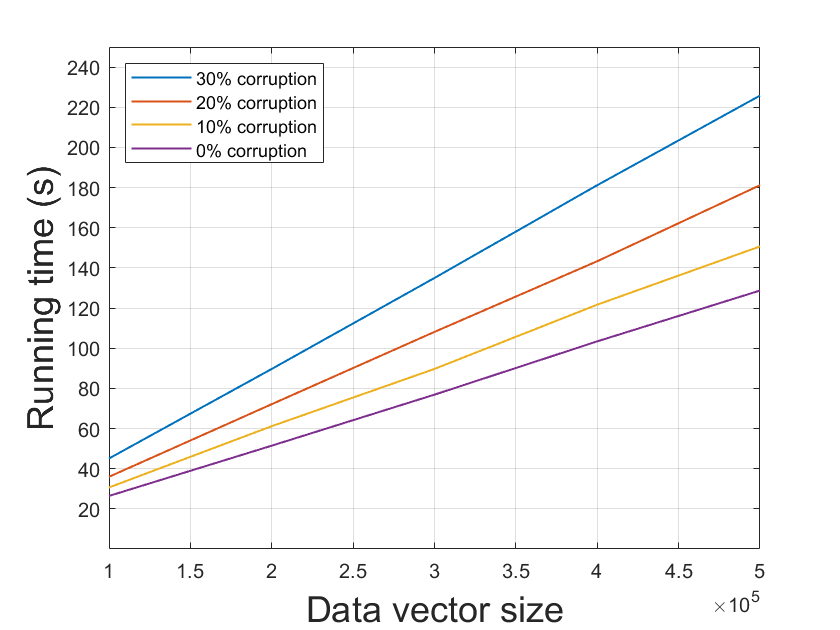}
}
\hspace{-0.2in}
\subfigure[When $\rho=20\%$] {
\label{fig:10}
\includegraphics[width=0.45\columnwidth]{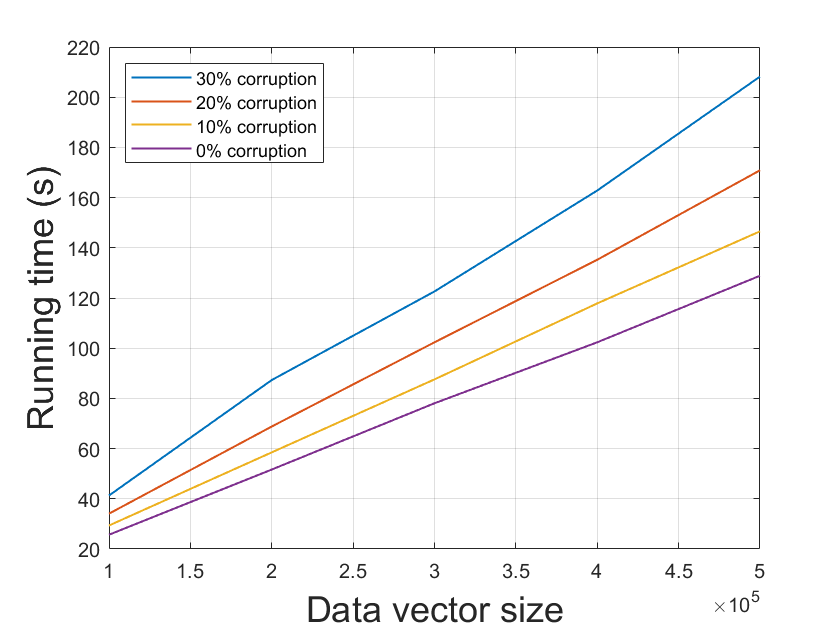}
}
\hspace{-0.2in}
\subfigure[When $\rho=10\%$] {
\label{fig:11}
\includegraphics[width=0.45\columnwidth]{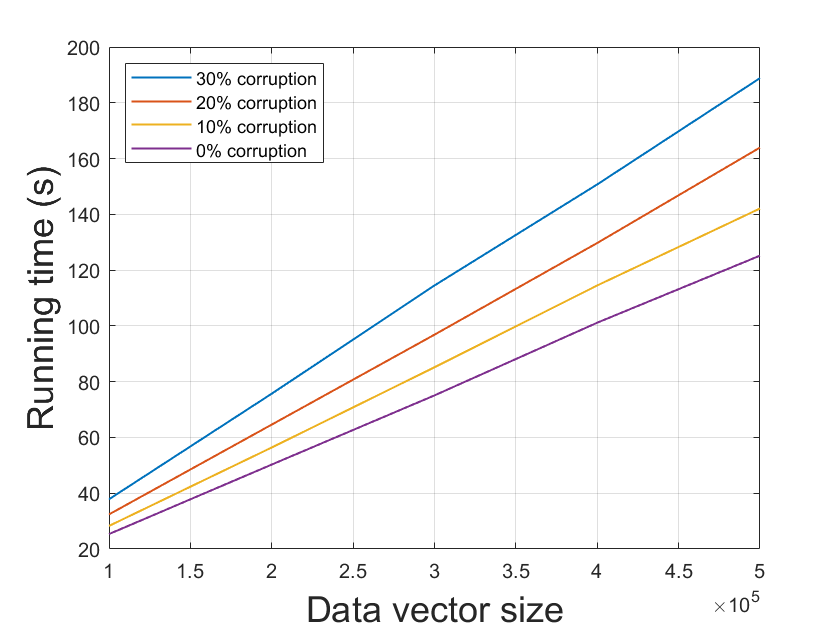}
}
\hspace{-0.2in}
\subfigure[When $\rho=0\%$] {
\label{fig:12}
\includegraphics[width=0.45\columnwidth]{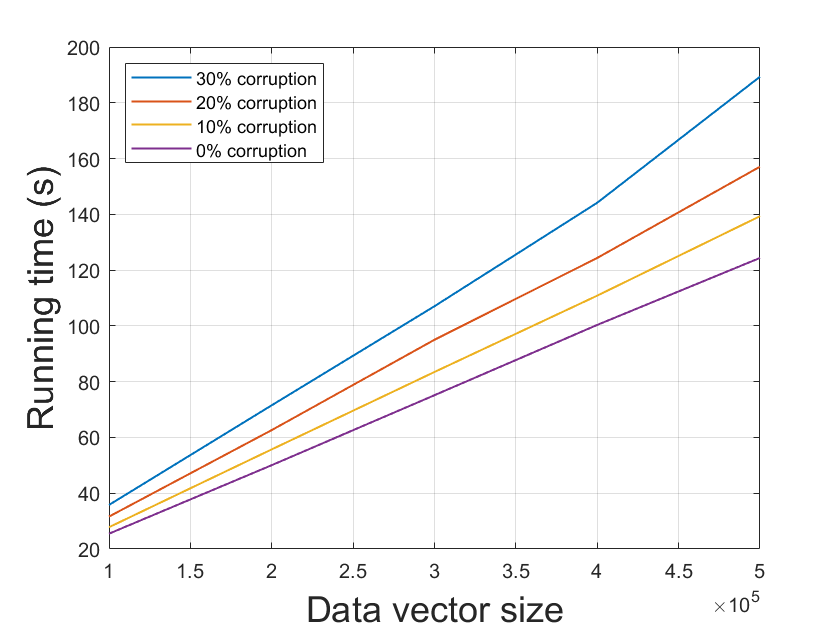}
}
\caption{Running time per client, where $n=500$.}
\label{fig9-12}
\end{figure}

\begin{figure}
\centering
\subfigure[When $\rho=30\%$] {
 \label{fig:13}
\includegraphics[width=0.45\columnwidth]{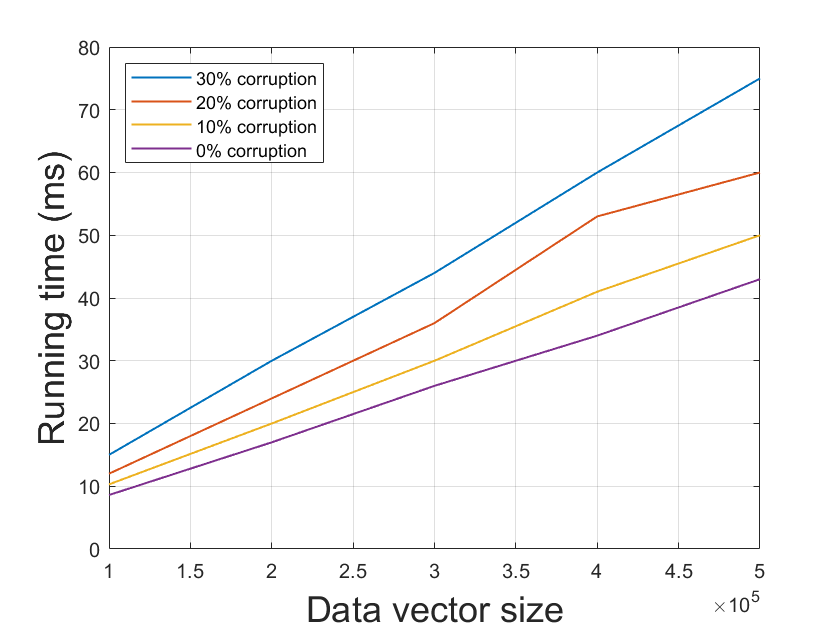}
}
\hspace{-0.2in}
\subfigure[When $\rho=20\%$] {
\label{fig:14}
\includegraphics[width=0.45\columnwidth]{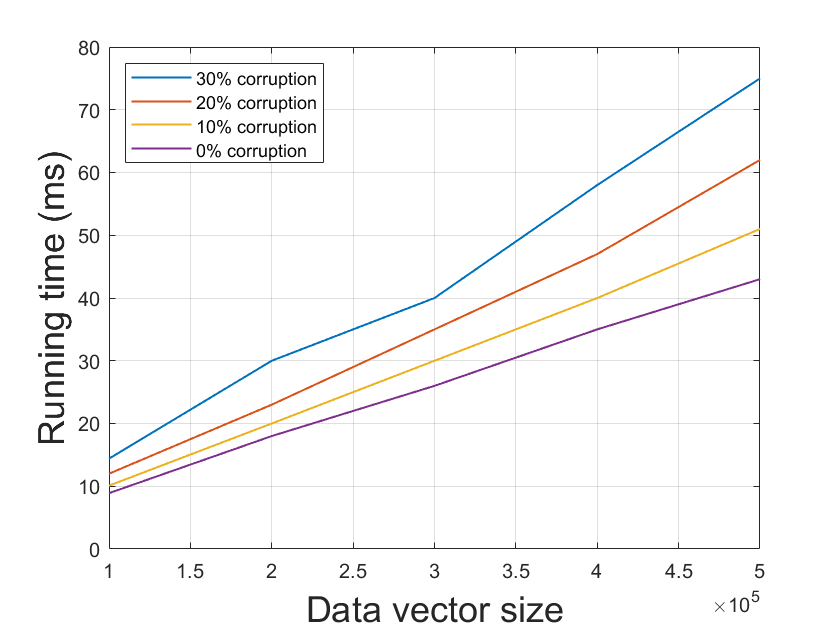}
}
\hspace{-0.2in}
\subfigure[When $\rho=10\%$] {
\label{fig:15}
\includegraphics[width=0.45\columnwidth]{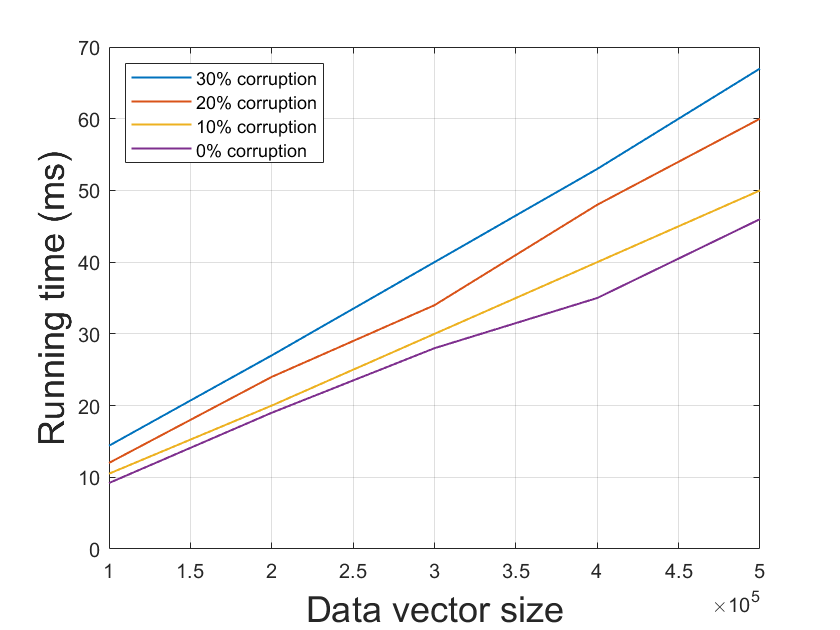}
}
\hspace{-0.2in}
\subfigure[When $\rho=0\%$] {
\label{fig:16}
\includegraphics[width=0.45\columnwidth]{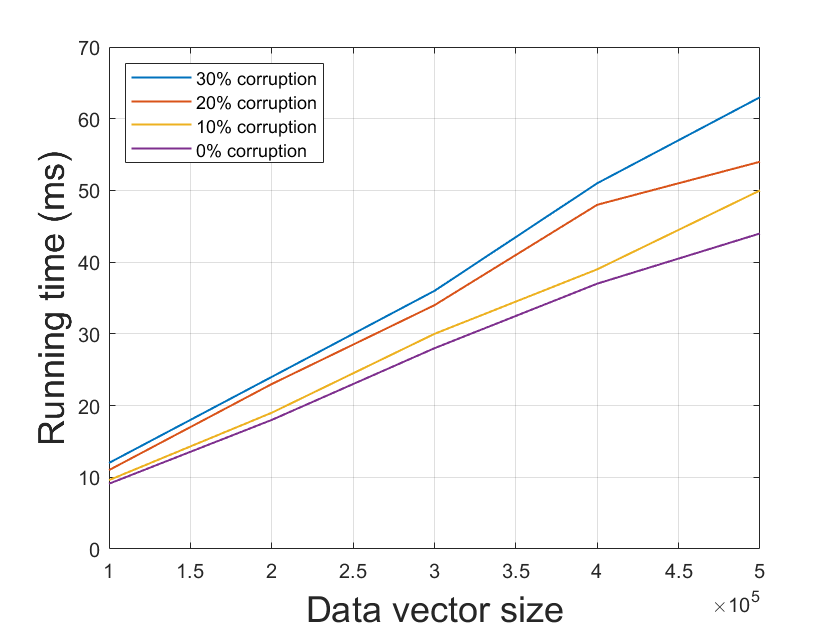}
}
\caption{Running time of the server, where $n=500$.}
\label{fig13-16}
\end{figure}

\noindent \textbf{Computation Overhead.} Fig. \ref{fig9-12} and Fig. \ref{fig13-16} depict the running times of each client and the server respectively when $n=500$.   As seen in Fig. \ref{fig9-12} and Fig. \ref{fig13-16}, for the fixed $\rho$ and $n$,  the running times of both the client and the server increase (approximately) linearly with the data vector size and increase significantly with $\gamma$.  In addition, Fig. \ref{fig5-8} and Fig. \ref{fig13-16} show that the data vector size has a greater impact on the computation cost of the server than the number of clients; see the reason described in Case 1.  Note that the curves in Fig. \ref{fig13-16} are unstable for the same reasons as in \ref{fig1-4}; see remark \ref{running}.
\vspace{0.2cm}

\begin{figure}
\centering
\subfigure[When $\rho=30\%$] {
 \label{fig:-5}
\includegraphics[width=0.45\columnwidth]{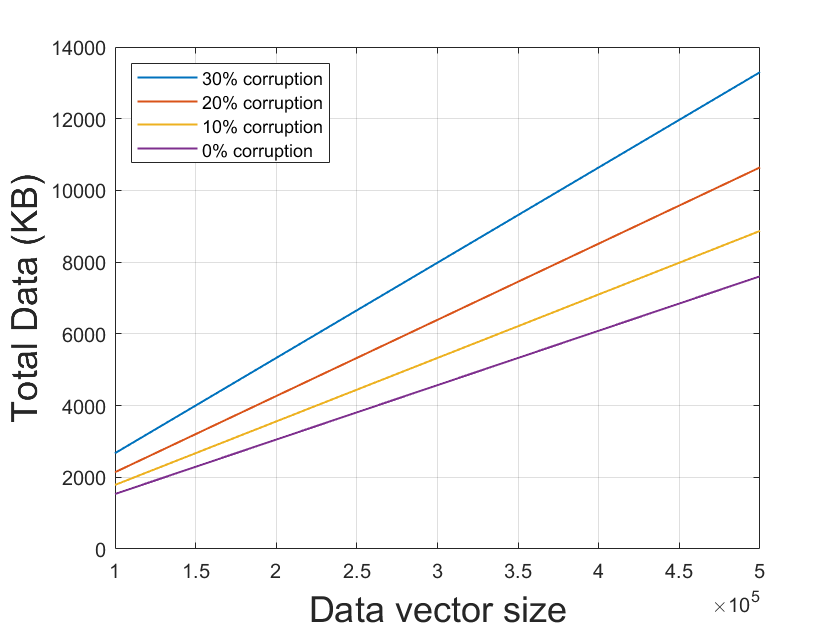}
}
\hspace{-0.2in}
\subfigure[When $\rho=20\%$] {
\label{fig:-6}
\includegraphics[width=0.45\columnwidth]{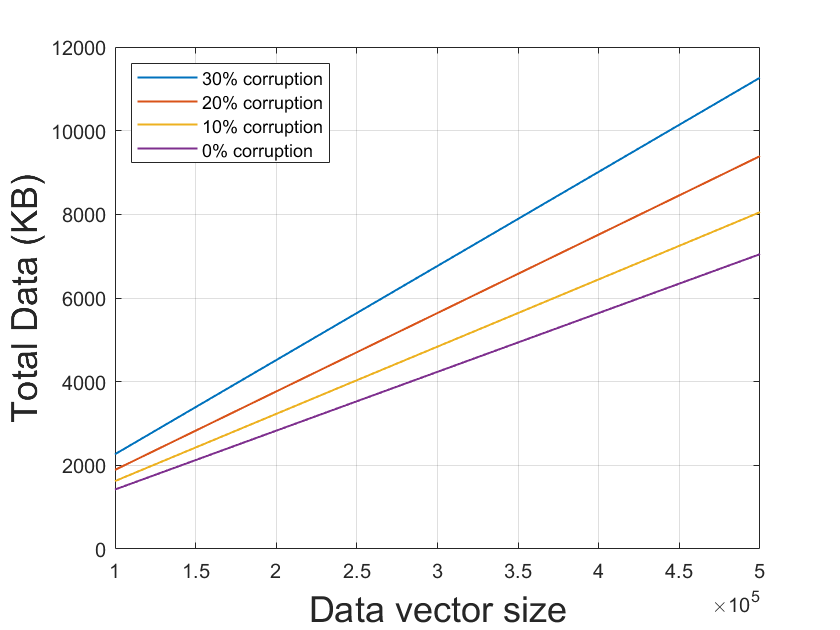}
}
\hspace{-0.2in}
\subfigure[When $\rho=10\%$] {
\label{fig:-7}
\includegraphics[width=0.45\columnwidth]{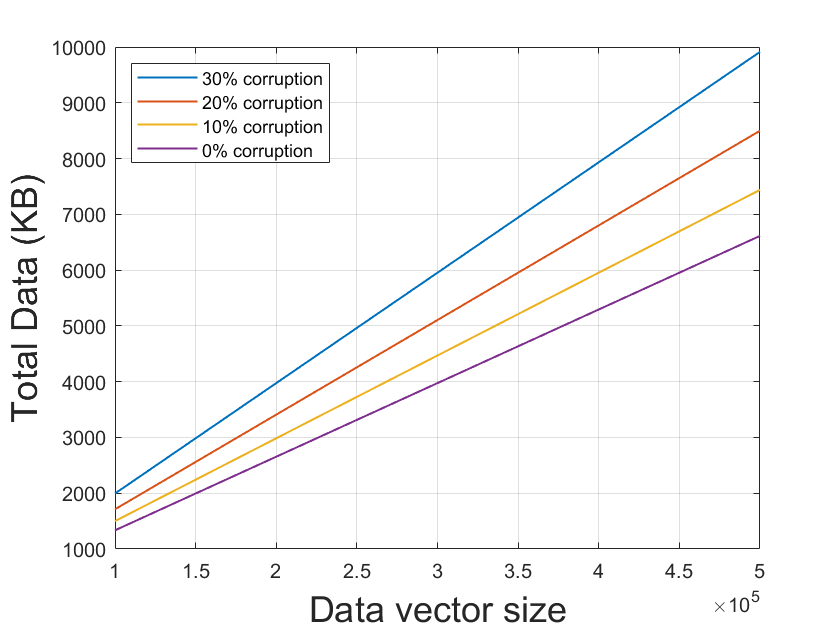}
}
\hspace{-0.2in}
\subfigure[When $\rho=0\%$] {
\label{fig:-8}
\includegraphics[width=0.45\columnwidth]{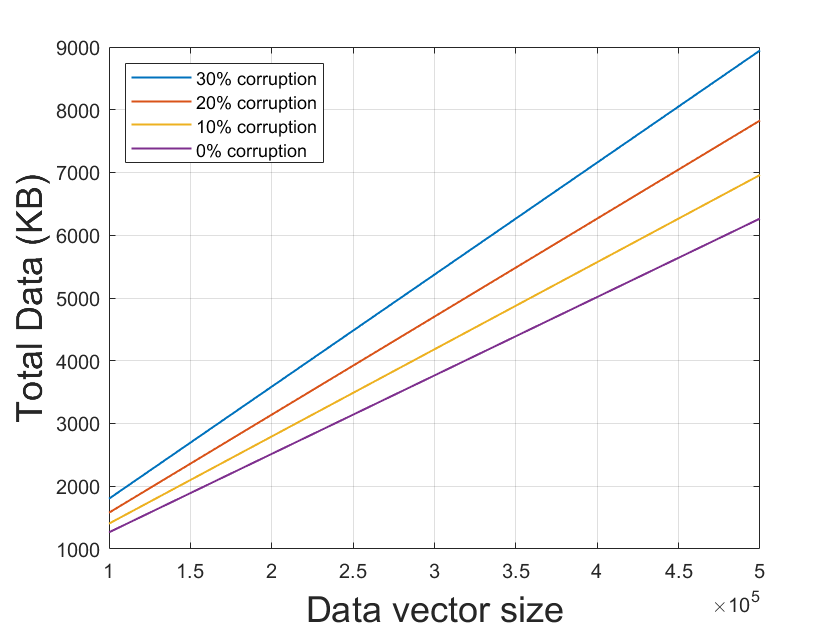}
}
\caption{Total data transfer per client, where $n=500$.}
\label{fig-5--8}
\end{figure}

\noindent \textbf{Communication Overhead.} Fig. \ref{fig-5--8} shows the total amount of data transferred by each client.  We can see that, for the fixed $\rho$ and $n$, the communication cost of each client increases linearly with the data vector size and increases significantly with $\gamma$.

\subsection{Performance Analysis under Case 3}
\begin{figure}
\centering
\subfigure[When $\gamma=30\%$] {
 \label{fig:17}
\includegraphics[width=0.45\columnwidth]{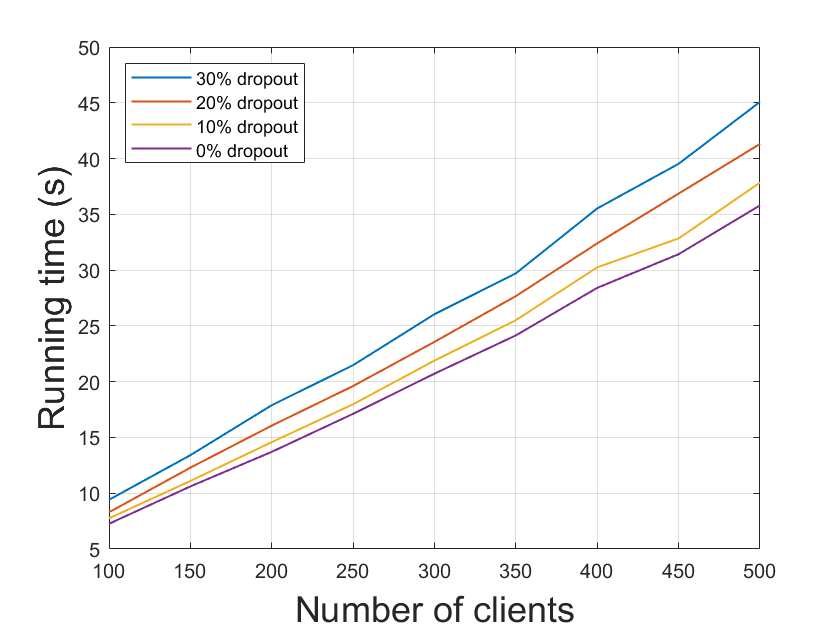}
}
\hspace{-0.2in}
\subfigure[When $\gamma=20\%$] {
\label{fig:18}
\includegraphics[width=0.45\columnwidth]{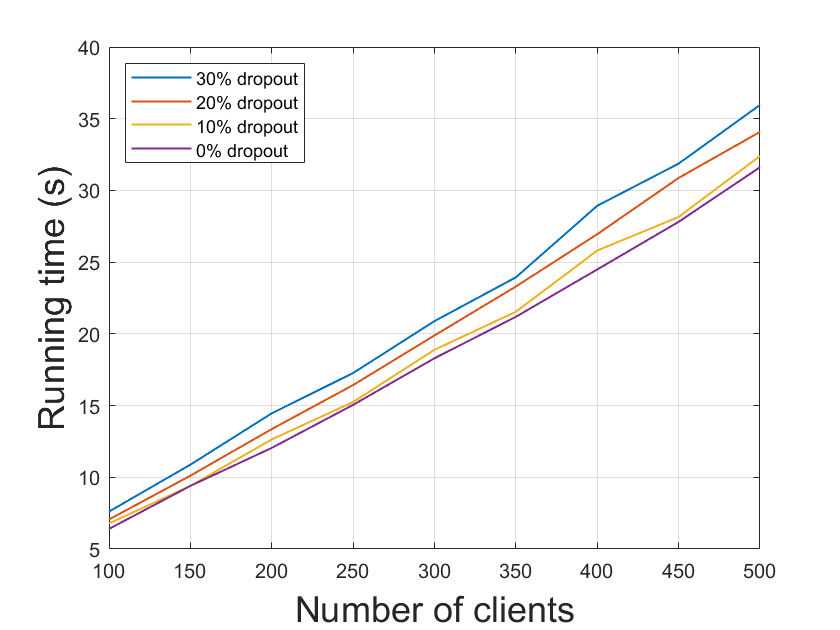}
}
\hspace{-0.2in}
\subfigure[When $\gamma=10\%$] {
\label{fig:19}
\includegraphics[width=0.45\columnwidth]{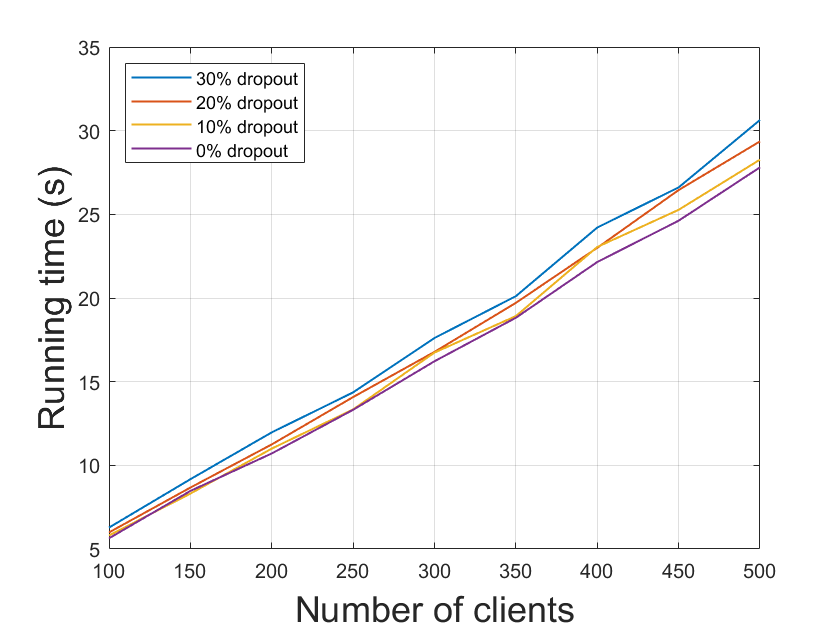}
}
\hspace{-0.2in}
\subfigure[When $\gamma=0\%$] {
\label{fig:20}
\includegraphics[width=0.45\columnwidth]{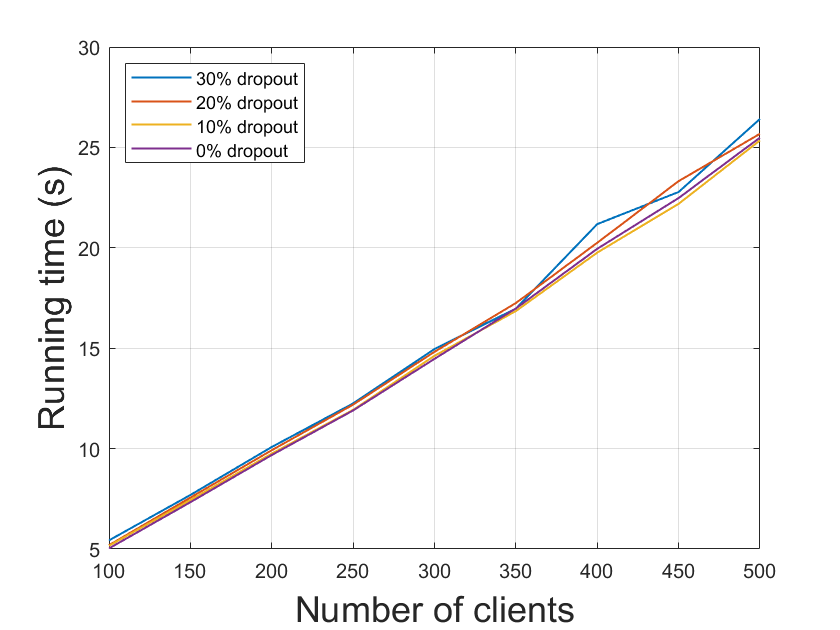}
}
\caption{Running time per client, where the data vector size is fixed to 100K.}
\label{fig17-20}
\end{figure}

\noindent \textbf{Computation Overhead.} Fig. \ref{fig17-20} depicts the running time of each client for different $\gamma$.   As seen in Fig. \ref{fig17-20}, for the fixed $\gamma$ and the data vector size, the running time of each client increase (approximately) linearly with $n$.  Besides, as $\gamma$ becomes smaller, the influence of $\rho$ on the client's computational cost also becomes smaller.  We omit the running time plot for the server because we can see from Fig. \ref{fig5-8} that the running time of the server remains almost unchanged with $n$ and does not change significantly with $\rho$ for the fixed data vector size and $\gamma$.  Note that the curves in Fig. \ref{fig17-20} are unstable for the same reasons as in \ref{fig1-4}; see remark \ref{running}. \vspace{0.2cm}

\noindent \textbf{Communication Overhead.}  We omit the data transfer plot for the client because it is not hard to see from Fig. \ref{fig-1--4} that the communication cost of each client remains almost unchanged with $n$ but increases significantly with $\rho$ for the fixed data vector size and $\gamma$.

\subsection{Performance Analysis under Case 4}
\begin{figure}
\centering
\subfigure[When $\gamma=30\%$] {
 \label{fig:21}
\includegraphics[width=0.45\columnwidth]{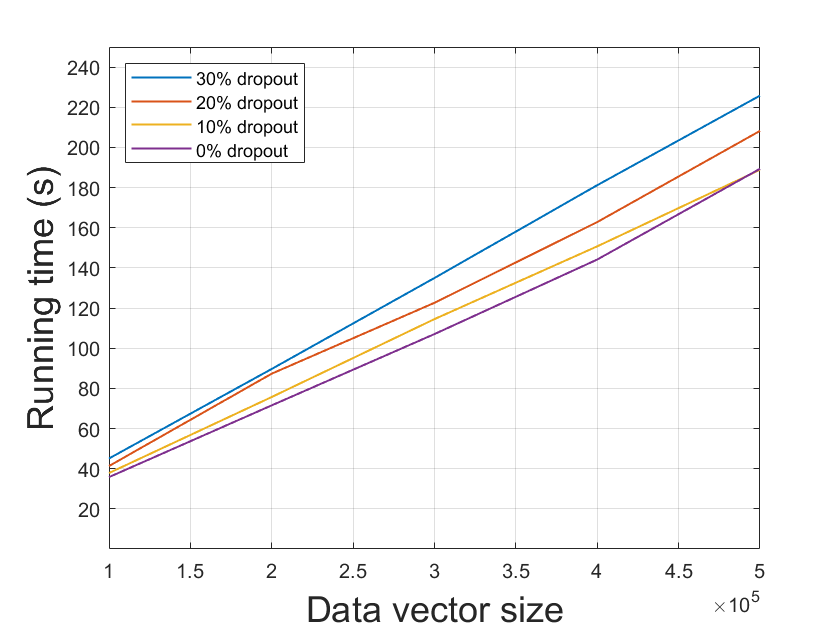}
}
\hspace{-0.2in}
\subfigure[When $\gamma=20\%$] {
\label{fig:22}
\includegraphics[width=0.45\columnwidth]{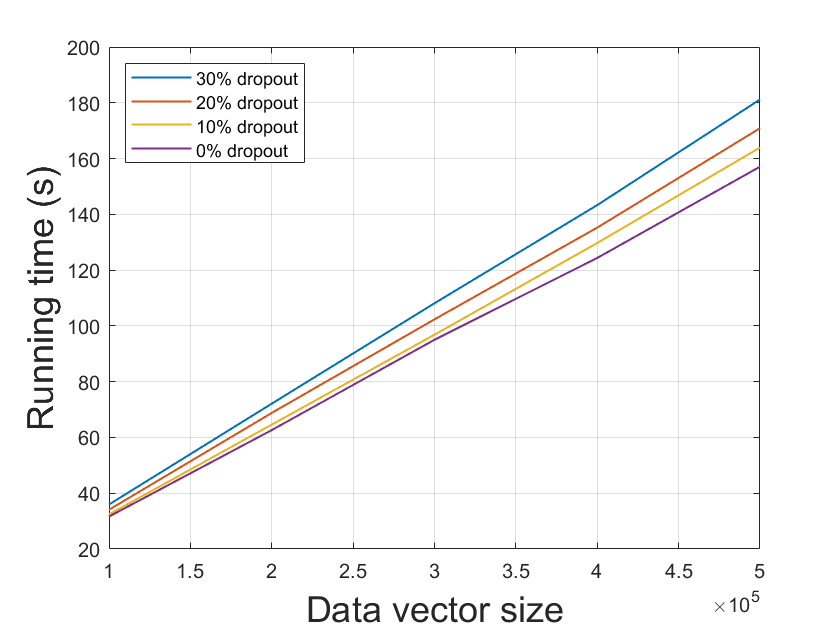}
}
\hspace{-0.2in}
\subfigure[When $\gamma=10\%$] {
\label{fig:23}
\includegraphics[width=0.45\columnwidth]{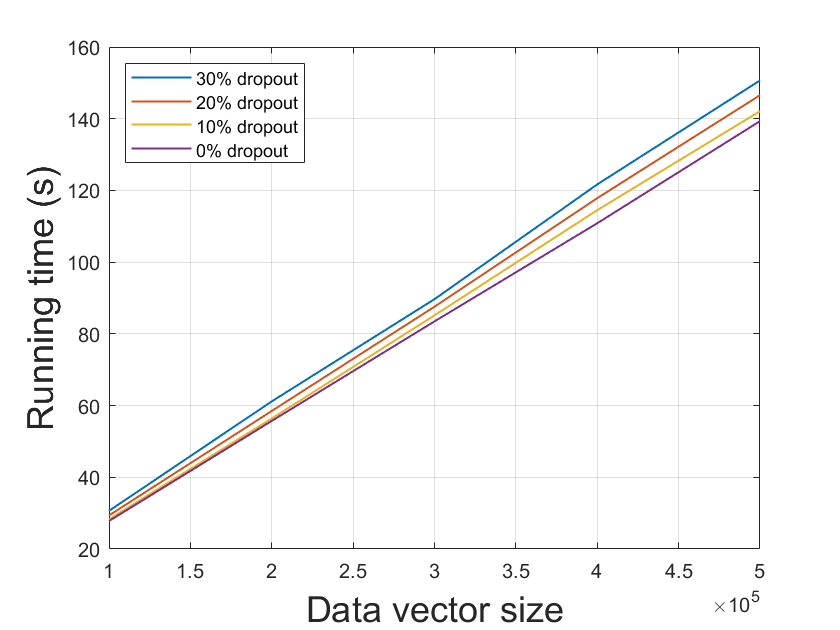}
}
\hspace{-0.2in}
\subfigure[When $\gamma=0\%$] {
\label{fig:24}
\includegraphics[width=0.45\columnwidth]{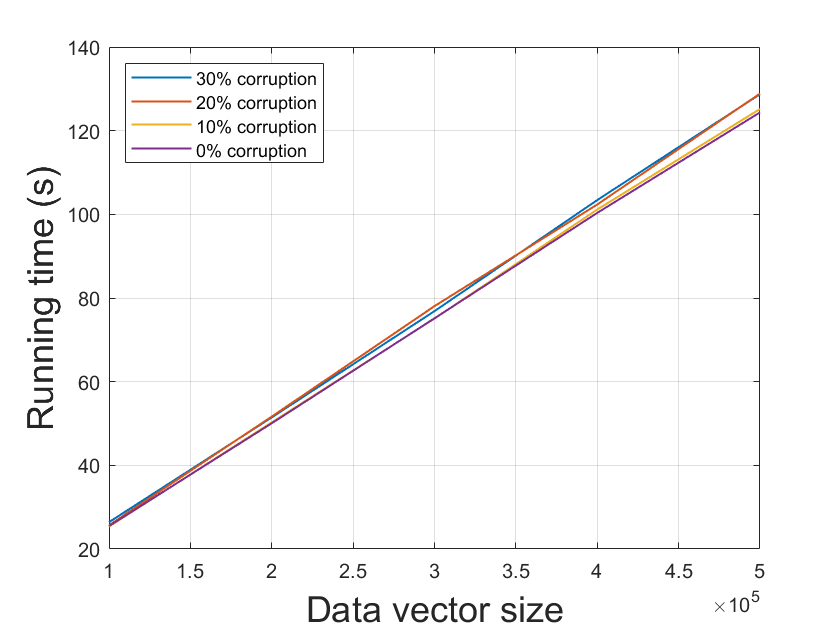}
}
\caption{Running time per client, where $n=500$.}
\label{fig21-24}
\end{figure}

\begin{figure}
\centering
\subfigure[When $\gamma=30\%$] {
 \label{fig:25}
\includegraphics[width=0.45\columnwidth]{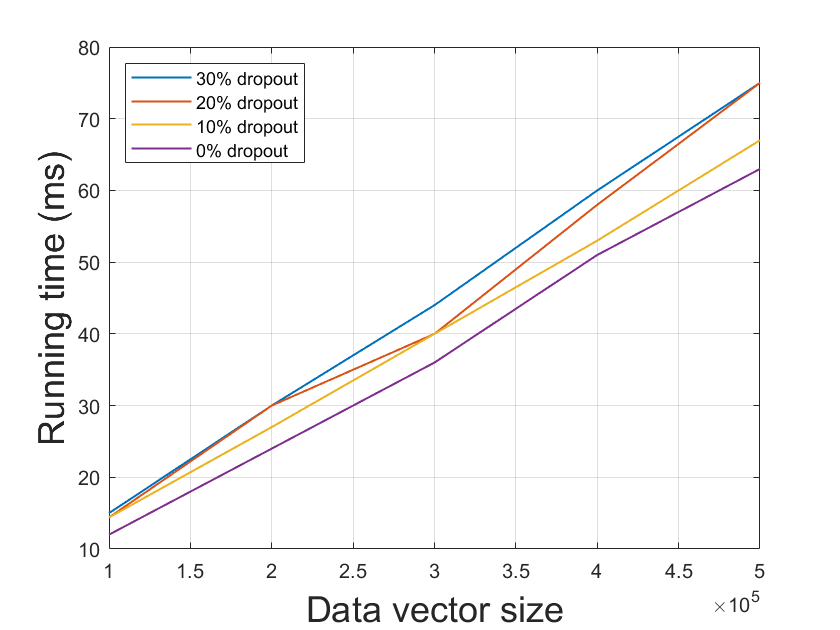}
}
\hspace{-0.2in}
\subfigure[When $\gamma=20\%$] {
\label{fig:14}
\includegraphics[width=0.45\columnwidth]{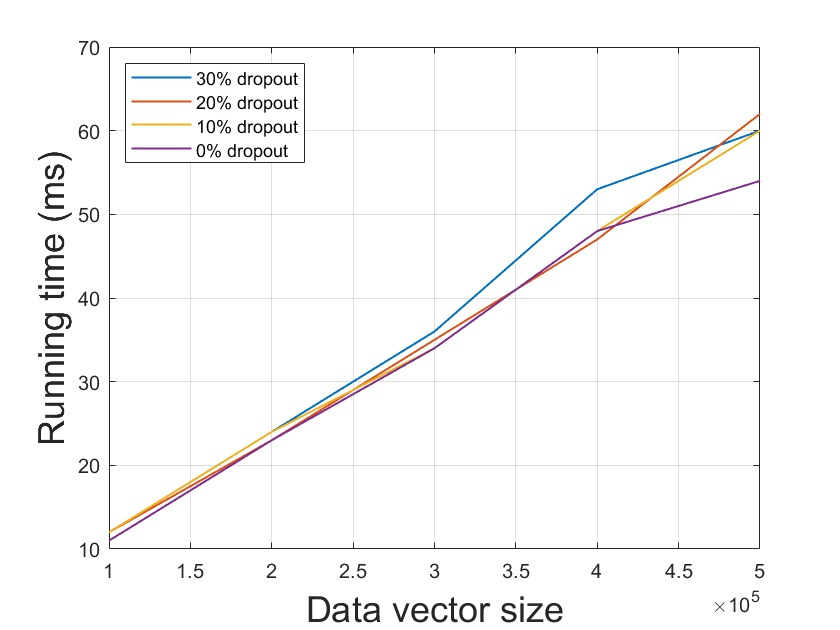}
}
\hspace{-0.2in}
\subfigure[When $\gamma=10\%$] {
\label{fig:15}
\includegraphics[width=0.45\columnwidth]{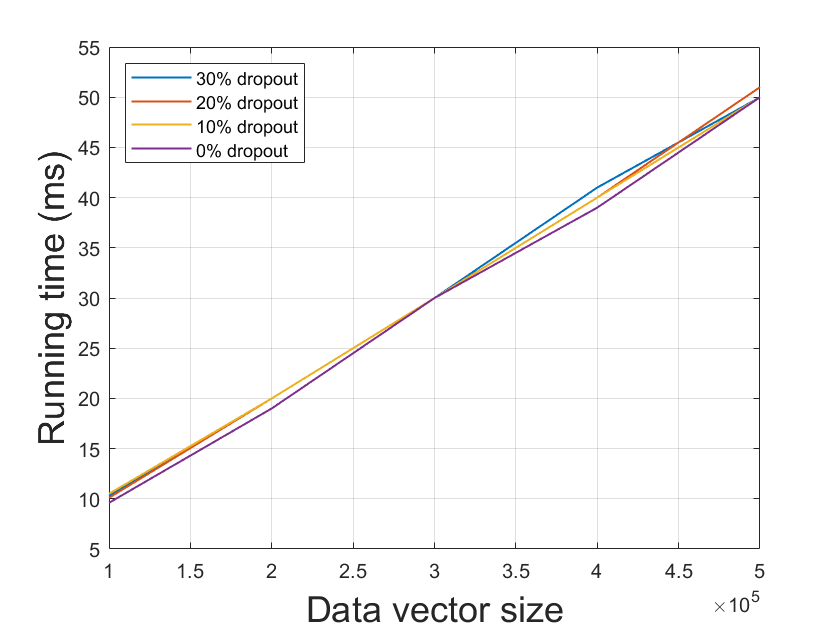}
}
\hspace{-0.2in}
\subfigure[When $\gamma=0\%$] {
\label{fig:16}
\includegraphics[width=0.45\columnwidth]{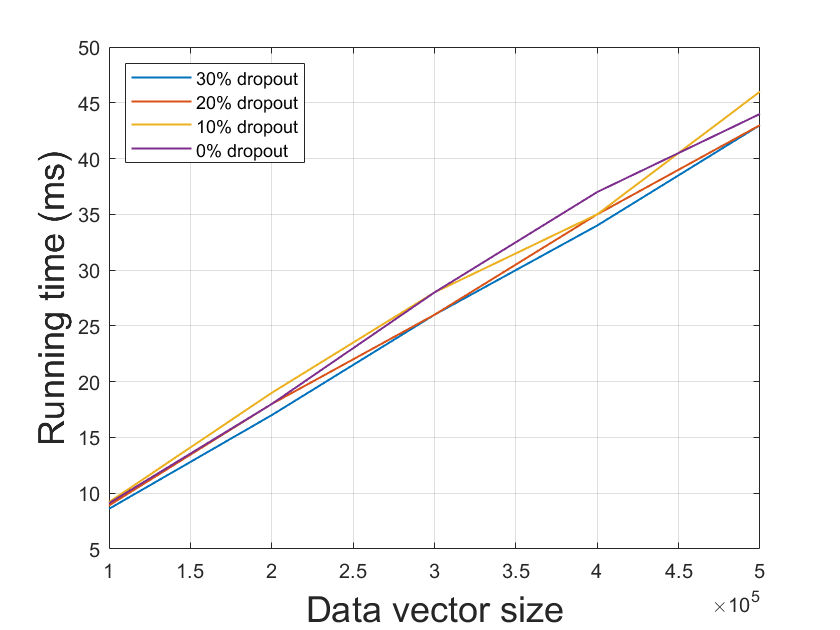}
}
\caption{Running time of the server, where $n=500$.}
\label{fig25-28}
\end{figure}

\noindent \textbf{Computation Overhead.} Fig. \ref{fig21-24} and Fig. \ref{fig25-28} depict the running times of each client and the server respectively when $n=500$.   As seen in Fig. \ref{fig21-24} and Fig. \ref{fig25-28}, for the fixed $\gamma$ and $n$,  the running times of both the client and the server increase (approximately) linearly with the data vector size.  Besides, as $\gamma$ becomes smaller, the impact of $\rho$ on the computational costs of both the client and the server also becomes smaller.  Note that the curves in Fig. \ref{fig25-28} are unstable for the same reasons as in \ref{fig1-4}; see remark \ref{running}. \vspace{0.2cm}

\noindent \textbf{Communication Overhead.} We omit the data transfer plot for the client because it is not hard to see from Fig. \ref{fig-5--8} that the communication cost of each client increases linearly
with the data vector size and increases significantly with $\rho$ for the fixed $n$ and $\gamma$.  Besides, as $\gamma$ becomes smaller, the impact of $\rho$ on the communication cost of each client also becomes smaller.\vspace{0.2cm}

\noindent \textbf{Conclusion.} Combing the above 4 cases, we conclude that 1) the computation cost of each client increases (approximately) linearly with both the number of clients and the data vector size for the fixed $\gamma$ and $\rho$; 2) the computation cost of the server increases (approximately) linearly with the data vector size and remains almost unchanged with $n$ for the fixed $\gamma$ and $\rho$; 3) $\gamma$ has a greater impact on the computation and communication overhead of each client and the server than $\rho$; and 4) the communication cost of each client remains almost unchanged with $n$, increases linearly with the data vector size, and increases significantly with both $\gamma$ and $\rho$.

\subsection{Comparison with Prior Secure Aggregation \cite{BIK17}}
\begin{figure*}
\centering
\subfigure[When $\rho,\gamma=30\%$] {
 \label{fig:CS1}
\includegraphics[width=0.5\columnwidth]{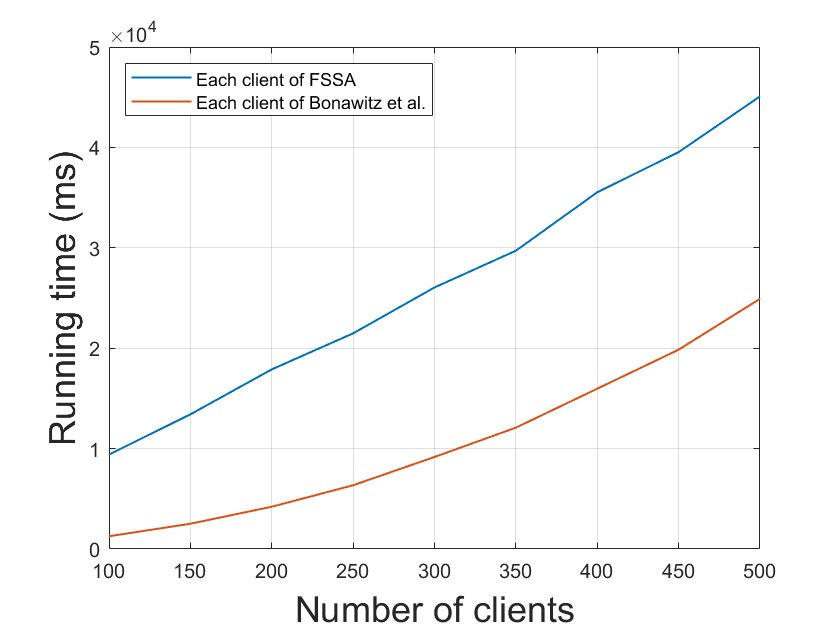}
}
\centering
\subfigure[When $\rho,\gamma=30\%$] {
\label{fig:GS2}
\includegraphics[width=0.5\columnwidth]{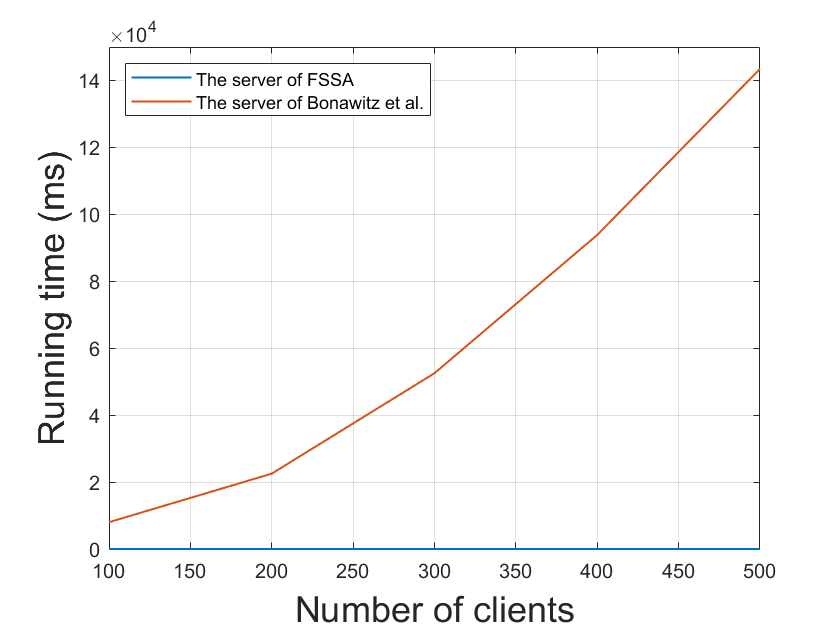}
}
\centering
\subfigure[When $\rho,\gamma=30\%$] {
\label{fig:CS1-2}
\includegraphics[width=0.5\columnwidth]{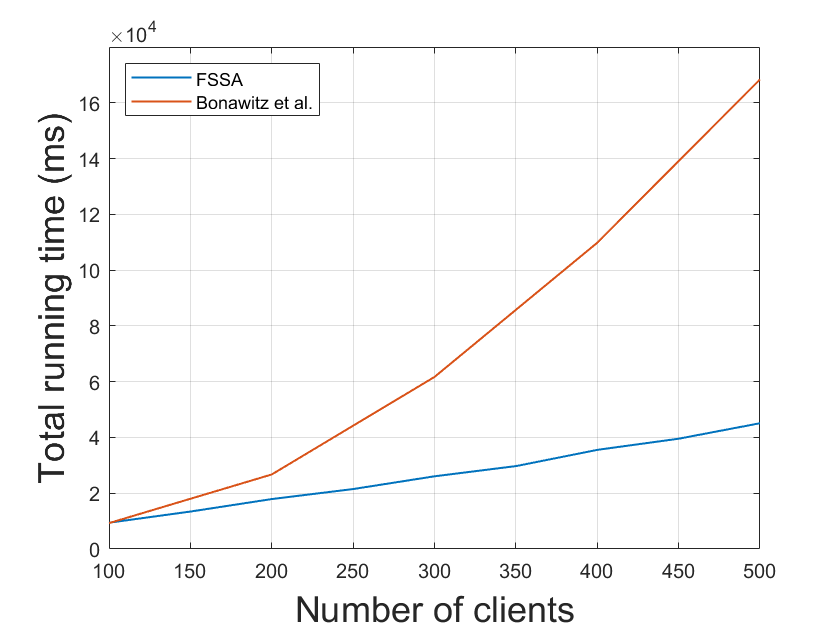}
}

\centering
\subfigure[When $\rho,\gamma=30\%$] {
\label{fig:GS3}
\includegraphics[width=0.5\columnwidth]{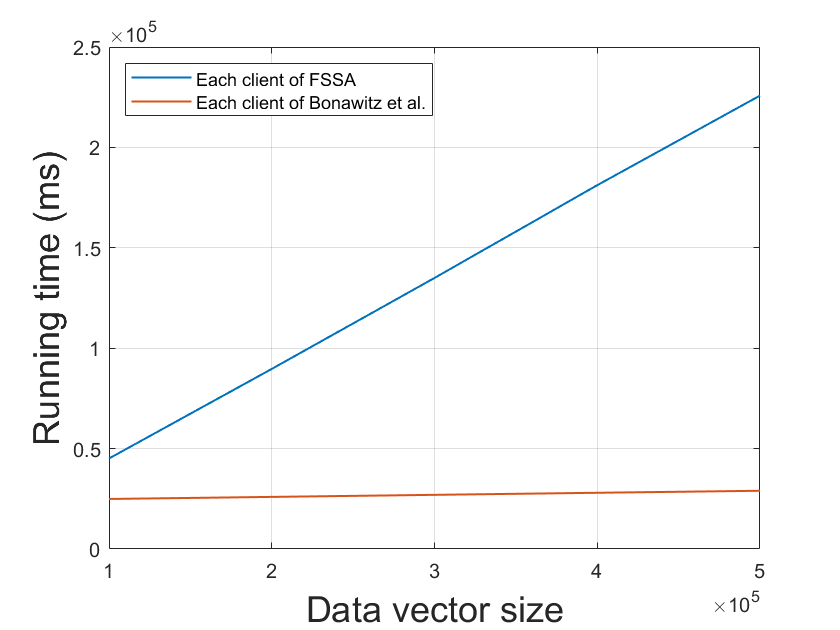}
}
\centering
\subfigure[When $\rho,\gamma=30\%$] {
\label{fig:GS4}
\includegraphics[width=0.5\columnwidth]{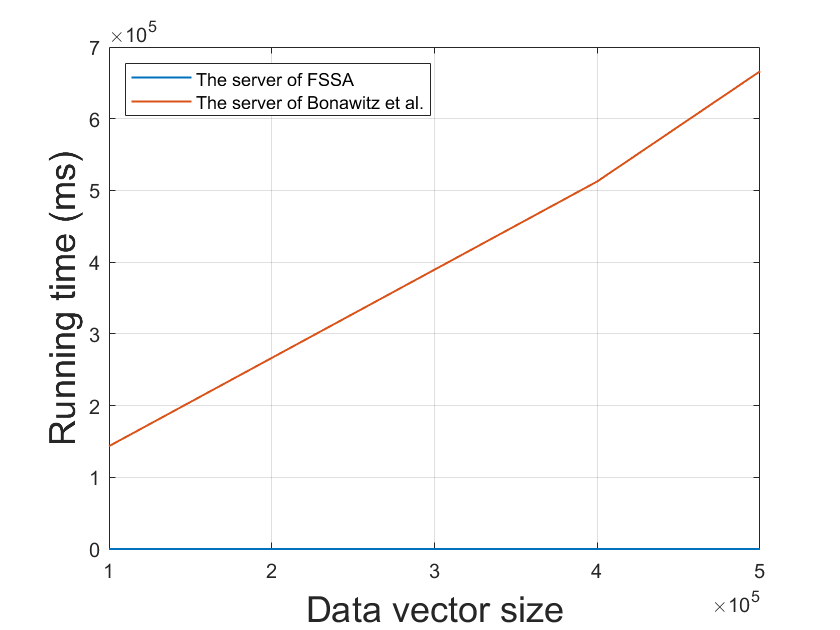}
}
\centering
\subfigure[When $\rho,\gamma=30\%$] {
\label{fig:GS3-4}
\includegraphics[width=0.5\columnwidth]{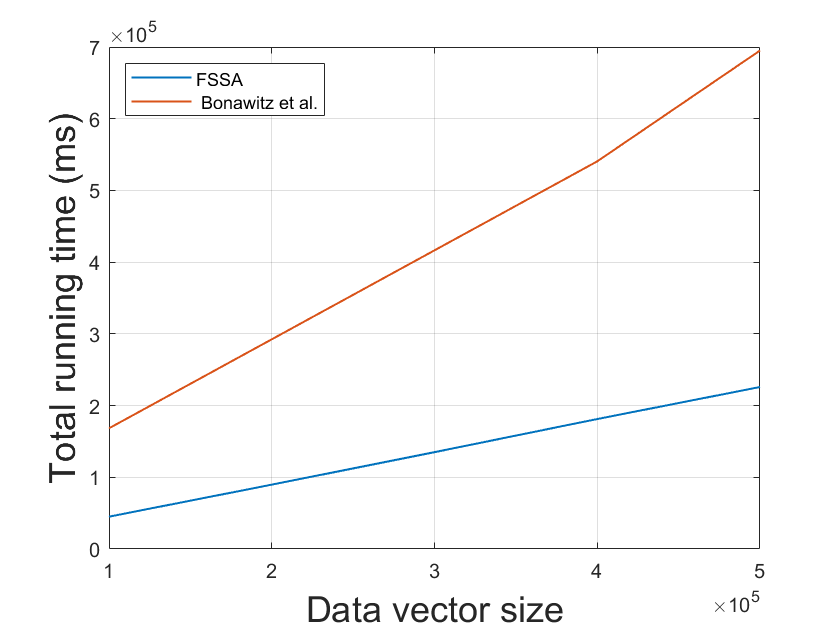}
}
\caption{Comparison between FSSA and the secure aggregation in \cite{BIK17} in terms of computational cost.
 (a) Computational cost of each client, where the data vector size is fixed to 100K. (b) Computational cost of the server, where the data vector size is fixed to 100K. (c) Total computation cost, where the data vector size is fixed to 100K.
 (d) Computational cost of each client, where $n=500$. (e) Computational cost of the server, where $n=500$. (f) Total computation cost, where $n=500$.  }
\label{fig-cs-comp}
\end{figure*}
As mentioned earlier, the secure aggregation protocol of \cite{BIK17} requires 4 rounds of communication between each client and the server in the honest-but-curious setting.  By contrast, our FSSA protocol only requires 3 rounds of communication between each client and the server in the honest-but-curious setting.   This means that our FSSA protocol has a lower dropout rate $\rho$ (recall that in FL, the more round, the higher the dropout rate) and thus achieves a higher model accuracy than the secure aggregation protocol in \cite{BIK17} when applied to FL.

Recall that the secure aggregation protocol of \cite{BIK17} used a Pseudorandom Generator (PRG) to reduce the communication.  However, as stated in \cite{BIK17}, the computational cost of PRG is much higher than that of key agreement, secret sharing\slash reconstruction, and authenticated encryption\slash decryption.  By contrast, FSSA does not require the expensive PRG operations, and its computational cost is largely determined by secret sharing.   However, we note that the data vectors shared using secret sharing in FSSA are longer than that in \cite{BIK17}.

We compare the computational cost of each client and the server of FSSA with the fixed $(\rho=30\%,\gamma=30\%)$ with that of \cite{BIK17} with the fixed $\rho=30\%$.  Similarly, we compare the communication cost of each client of FSSA with the fixed $(\rho=30\%,\gamma=30\%)$ with that of \cite{BIK17} with the fixed $\rho=30\%$.  As depicted in Fig. \ref{fig-cs-comp},  compared with \cite{BIK17}, FSSA has a higher computational cost per client, a lower computational cost on the server, and a lower total computational cost (i.e., sum of the computation costs of each client and the server).  Therefore, we conclude that FSSA is more computationally efficient than  \cite{BIK17},  especially as the number of clients increases.

\begin{figure}
\centering
\subfigure[When $\rho,\gamma=30\%$] {
 \label{fig:comu}
\includegraphics[width=0.45\columnwidth]{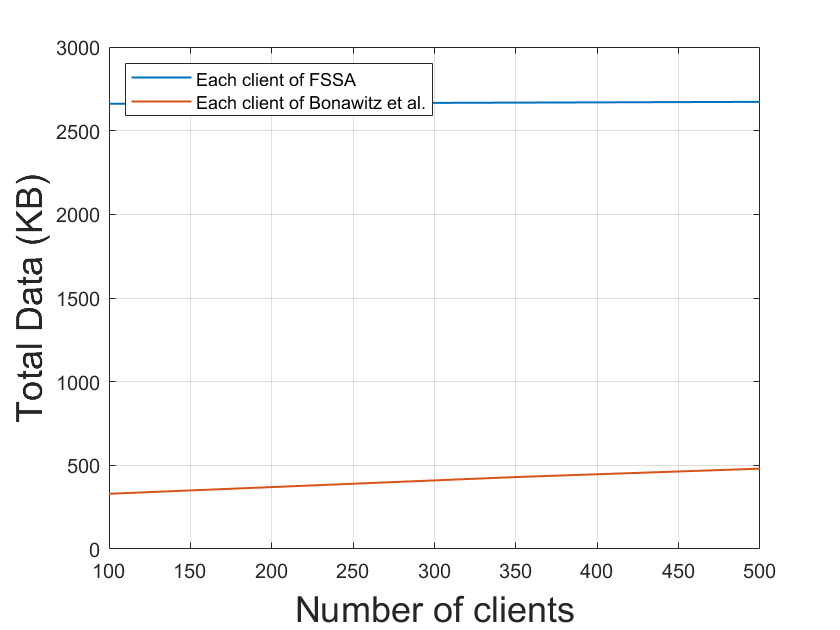}
}
\hspace{-0.2in}
\subfigure[When $\rho,\gamma=30\%$] {
\label{fig:comu1}
\includegraphics[width=0.45\columnwidth]{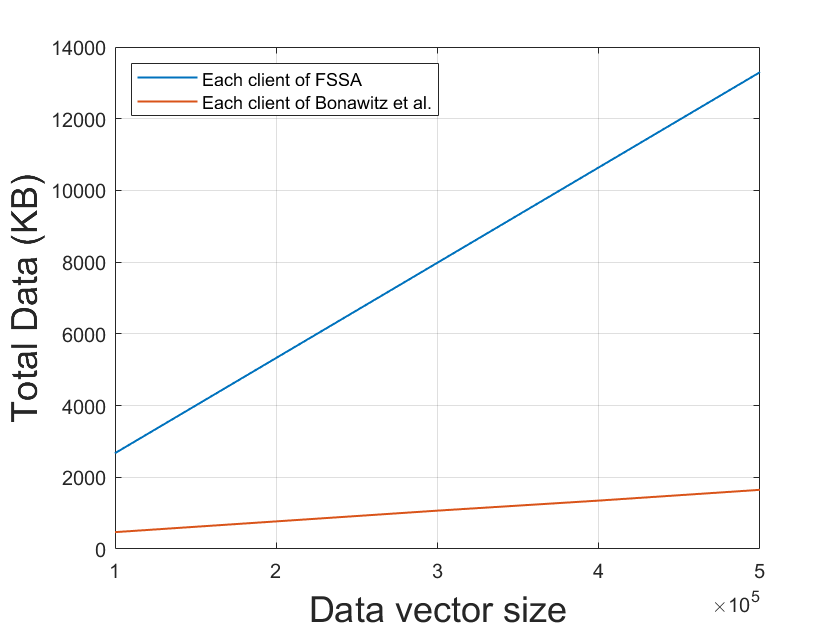}
}
\caption{Comparison between FSSA and the secure aggregation in \cite{BIK17} in terms of (a) total data transfer per client, where the data vector size is fixed to 100K, and (b) total data transfer per client, where $n=500$. }
\label{fig-comu}
\end{figure}

With respect to communication overhead, FSSA performs worse than \cite{BIK17}, as shown in Fig. \ref{fig-comu}.   However, in the real-world applications of FL, it is not uncommon for each client to transmit large chunks of data (i.e., megabytes) in each iteration of the model training.  Therefore, compared with \cite{BIK17}, FSSA incurs larger total communication cost, which usually clients can accommodate in most practical applications.  Nevertheless, FSSA requires fewer communication rounds, which achieves a lower dropout rate without incurring any extra computational cost.

\section{Conclusion}
\label{Sec:7}
In this paper, we investigated how to reduce the number of communication rounds in a secure aggregation protocol, and proposed an efficient 3-round secure aggregation protocol named FSSA.   FSSA is resilient against client dropouts and its computation and communication overhead is low enough to be used in mobile applications.  In addition, compared with Bonawitz \textit{et al.}'s secure aggregation protocol, FSSA not only requires fewer communication rounds, but also has less total computational overhead.  However,  FSSA has a larger communication overhead than Bonawitz \textit{et al.}'s secure aggregation protocol.  We leave reducing the communication overhead of FSSA as a future work.

\end{document}